\documentclass[journal,twoside,web]{ieeecolor}
\usepackage{generic}
\usepackage{cite}
\usepackage{mathtools, cuted}
\usepackage{amsmath,amssymb,amsfonts}
\usepackage{graphicx} 
\graphicspath{ {./Figures/} }
\usepackage[figurename=Fig.]{caption}
\usepackage{moreverb}
\usepackage[ansinew]{inputenc}
\usepackage{psfrag}

\usepackage{amsmath}
\usepackage{amsthm}
\theoremstyle{remark}
\newtheorem{assumption}{Assumption}
\theoremstyle{remark}
\newtheorem{theorem}{Theorem}[section]
\theoremstyle{remark}
\newtheorem{remark}{Remark}
\theoremstyle{remark}
\newtheorem{definition}{Definition}
\theoremstyle{remark}
\newtheorem{problem}{Problem}
\theoremstyle{remark}

\usepackage{amssymb}
\usepackage{float}
\usepackage{multirow}
\usepackage{lineno}
\usepackage{setspace}
\usepackage{flushend}
\usepackage{array}
\usepackage{tabu}
\usepackage{bm}
\usepackage{color}
\usepackage{url}
\usepackage{dsfont}
\usepackage[figuresright]{rotating}
\usepackage{subfigure}
\usepackage{svgcolor}
\usepackage{makecell}
\usepackage{booktabs}
\usepackage{textcomp}
\usepackage{scalerel}
\usepackage{eurosym}
\usepackage{braket}
\usepackage{xcolor}
\newcommand\BibTeX{{\rmfamily B\kern-.05em \textsc{i\kern-.025em b}\kern-.08em
		T\kern-.1667em\lower.7ex\hbox{E}\kern-.125emX}}
\def\tsc#1{\csdef{#1}{\textsc{\lowercase{#1}}\xspace}}
\tsc{ASV}
\tsc{USV}
\tsc{ASC}
\tsc{PFC}
\tsc{GPS}
\tsc{IMU}
\tsc{SBC}
\tsc{ESO}
\tsc{ISS}

\def\BibTeX{{\rm B\kern-.05em{\sc i\kern-.025em b}\kern-.08em
    T\kern-.1667em\lower.7ex\hbox{E}\kern-.125emX}}
\begin{document}
\title{Velocity and Disturbance Robust Non-linear Estimator for Autonomous Surface Vehicles with Reduced Sensing Capabilities}
\author{Guillermo~Bejarano,
        Sufiyan~N-Yo,
        and~Luis~Orihuela 
\thanks{The authors would like to acknowledge Junta de Andaluc{\'\i}a (project reference PY18-RE-0009) for funding this work.}        
\thanks{G. Bejarano, S. N-Yo, and L. Orihuela are with the Engineering Department, Universidad Loyola Andaluc{\'\i}a, Avenida de las Universidades s/n. 41704 Dos Hermanas, Sevilla, Espa\~na.  E-mail:  gbejarano@uloyola.es, sufiyanyo@uloyola.es, dorihuela@uloyola.es}
\thanks{Corresponding author: G. Bejarano (email: gbejarano@uloyola.es).}}
\maketitle

\begin{abstract}
This paper presents a robust non-linear state estimator for autonomous surface vehicles, where the movement is restricted to the horizontal plane. It is assumed that only the vehicle position and orientation can be measured, being the former affected by bounded noises. Then, under some fair standard assumptions concerning the maximum velocities and acceleration rates of the vehicle, the estimator is able to reconstruct not only the velocities, but also the lumped generalised disturbances, that cluster external disturbances, non-linearities, and unmodelled dynamics. The observer is easily tunable by the user, with a set of four scalars, two of them related to the velocity of convergence of the estimator, and the other two parameters to set the desired trade-off between noise sensitivity and disturbance rejection. Several simulations with a well-known test-bed craft are provided to show how the proposed algorithm outperforms previous ones in the literature.
\end{abstract}

\begin{IEEEkeywords}
Autonomous surface vehicles, disturbance estimation,  marine systems, noise sensitivity, velocity estimation.
\end{IEEEkeywords}

\section{Introduction}
\label{sec:introduction}
\IEEEPARstart{A}{utonomous} surface vehicles (ASVs), also known as unmanned surface vehicles (USVs) or autonomous surface crafts (ASCs), have drawn increased attention in the past several decades due to their ability to perform a wide variety of high-risk tasks in harsh or otherwise unreachable marine environments without human intervention. Moreover, they address data acquisition with lower cost and human risk, thus boosting ocean resource exploration \cite{liu2016unmanned}. Climate change, environmental monitoring, and national security issues such as border surveillance, among other civil and military applications, have led to a growing interest in both scientific and military communities for this kind of vehicles. In particular, their versatility and enhanced maneuverability in shallow waters (riverine and coastal regions) compared to larger crafts are highly appreciated in very diverse fields \cite{wynn2014autonomous,zhang2015future}.

To address such tasks, it is critical to design an efficient and robust motion control system, which allows ASVs to work in unstructured and unpredictable environments in a fully autonomous way. This raises a series of theoretical challenges that have caused ASV motion control to become a research hotspot in recent years \cite{miao2017compound,wang2019fuzzy,xia2019improved}. 


Effective and reliable motion control of ASVs is a challenging task, considering the complex model non-linearities, unmodelled hydrodynamics, internal parametric model uncertainties, and unmeasurable disturbances due to wind, waves, and currents, which may affect or even shatter the vessel control performance \cite{breivik2010topics}. To a large extent, many control strategies rely on how disturbances and uncertain terms are estimated, since they can be compensated if the estimation is accurate enough \cite{chen2017sliding,CUI201645,lekkas2014integral,liu2019state,skjetne2005adaptive,wang2019fuzzy,yin2017tracking,peng2020output}. In that case, the controller design is simplified, and the motion control performance is enhanced. Furthermore, efficient motion control of ASVs depends heavily on a suitable navigation system where sensing and state estimation are key factors. Despite the growing availability of more effective, compact, and affordable navigation equipment, including global and differential positioning systems (GPSs and DGPSs), as well as inertial measurement units (IMUs) \cite{manley2008unmanned}, in general and especially in implementations with a reduced number of sensors, only the position and orientation of the ASV are available \cite{liu2016unmanned}. Velocities (known as surge, sway, and yaw when considering the ASV to move in a horizontal plane) and accelerations must be usually recovered by using only the position-heading information, measured by a DGPS and a gyrocompass.

On the one hand, many works have addressed the problem of counteracting disturbances and model uncertainties under two main perspectives: a) model-based control techniques, where an adaptive control strategy is applied to estimate and then compensate disturbances and unmodelled dynamics \cite{chen2017sliding,CUI201645,lekkas2014integral,skjetne2005adaptive,dai2018adaptive,wang2019fuzzy,peng2019path}; and b) integral-based control techniques, where the effect of disturbances and unmodelled dynamics is attenuated by introducing an additional integral action in the controller \cite{do2006underactuated,feemster2011comprehensive,pan2013efficient}. Integral-based control techniques do not provide any detailed insight into ASV dynamics caused by disturbances, but ensure local stability and derive simpler controllers than model-based control techniques, where parametric adaptive methods \cite{skjetne2005adaptive}, robust control \cite{lekkas2014integral}, sliding-mode control \cite{chen2017sliding,CUI201645}, and fuzzy logic systems and neural networks have been applied to handle disturbances and uncertainties \cite{dai2018adaptive,wang2019fuzzy,peng2019path}. However, all these control strategies assume accurate velocity measurement, which may not be available in ASV implementations with only position sensors, due to technical reasons or just cost saving. 

On the other hand, concerning state estimation, high-resolution estimates of the velocities are often required to compute control actions. However, measurements provided by GPSs/DGPSs and gyrocompasses are usually influenced in practical applications by environmental noises. Thus, high-gain observers \cite{chen2012robust,tee2006control,ZHANG20111430} and different varieties of the well-known Kalman filter \cite{motwani2013interval,peng2009adaptive,tran2014tracking} have been applied to deal with state estimation. Peng \emph{et al.} propose an adaptive unscented Kalman filter (UKF) without \emph{a priori} knowledge of noise distribution \cite{peng2009adaptive}. Motwani \emph{et al.} propose a robust heading estimation strategy named
after an interval Kalman filter (IKF) in order to bound model uncertainties affecting the state estimation \cite{motwani2013interval}. Tran \emph{et al.} propose to estimate the position and velocity through an extended Kalman filter (EKF) based on a practical dynamic model and both GPS and compass measurements \cite{tran2014tracking}. However, in all cited works the state vector and the vessel uncertainties and disturbances cannot be simultaneously estimated.

Recently, Peng \emph{et al.} have addressed the simultaneous estimation of unmeasured velocities and uncertainties \cite{peng2016cooperative,peng2017distributed}, but persistent excitation of the vessel is required, while other works have applied extended state observers (ESO) to estimate both velocities and disturbances without persistence of exciting \cite{peng2018output,peng2019output,gu2019observer,zhang2019fixed,yiang2020line,liu2019state}. In particular, Liu \emph{et al.} have proposed a finite-time non-linear ESO where the convergence of the estimation of the disturbance terms does not depend on the control law \cite{liu2019state}, unlike other observer designs for non-linear systems where the uncertainty identification depends on the tracking error dynamics \cite{li2016adaptive,li2016hybrid}. Nevertheless, the ESO proposed in \cite{liu2019state} consider only noiseless position and orientation measurements. 

In this work, the observer structure proposed in \cite{liu2019state} is considered as a starting point, but several contributions are made and are listed below:
\begin{itemize}
	\item Noise in the position measurement is considered, which enhances the applicability of the observer to realistic applications. 
	\item The full state vector is divided into positional and rotational state vectors, for which two different observers are envisioned, thus providing additional degrees of freedom.
	\item Four meaningful scalar tuning parameters are considered, which allow to clearly define both the velocity of convergence (for the positional and rotational dynamics), and the user-defined trade-off between estimation quality and noise sensitivity, according to the expected noise level in the measured variables.
\end{itemize}

Concerning the first contribution, considering noise involves a challenge for the stability proof, since it is necessary to move from a standard stability condition, like the one developed in \cite{liu2019state}, to a robust one. Moreover, the dynamics of the positional estimation errors are intended to be input-to-state stable with respect to the measurement noises, in addition to the lumped disturbances, whereas the asymptotic estimation error bound is intended to be minimized. The design procedure is thus transformed from a feasibility problem to an optimization problem with Linear Matrix Inequalities (LMI) constraints. This step is not trivial, even more, when the objective is to keep a set of simple tuning parameters, with a direct physical meaning, that the user can tune to modify the performance of the estimator, as stated in the third contribution. This contrasts with the estimator in \cite{liu2019state}, where, despite proposing a more reduced set of parameters, the physical meaning of the latter is not clearly stated and the tuning procedure could be tricky in the case of noisy measurements.

The remainder of the work is organised as follows. Section \ref{secModProblemFormulation} presents the ASV model and formulates the estimation problem. The proposed observer design is detailed in Section \ref{secObserverDesign}, whereas Section \ref{secSimulation} provides a variety of simulation results showing how the capabilities of the ESO in \cite{liu2019state} are improved by the proposed observer, being also extended to the case where noisy positional measurements are considered. The tuning capabilities of the proposed observer are also illustrated. Finally, Section \ref{secConclusions} summarises the main conclusions and some future lines of research are proposed.

\section{Problem formulation} \label{secModProblemFormulation}
\subsection{Preliminaries} \label{subsecPreliminaries}

\begin{definition} \label{ISSDefinition} \cite{sontag2008input} Consider a non-linear system whose dynamics are shown in \eqref{eqsystemISS}:
	\begin{equation}
		\dot{\bm{x}}(t) = \bm{f}(\bm{x}(t),\bm{u}(t)),
		\label{eqsystemISS}
	\end{equation}
	where $\bm{x}(t) \in \mathbb{R}^n$ denotes the state vector, $\bm{u} \in \mathbb{R}^m$ is the input vector, and $\bm{f} : \mathbb{R}^n \times \mathbb{R}^m \rightarrow \mathbb{R}^n$ is the input-state map, for some positive integers $n$, $m$. The system defined in \eqref{eqsystemISS} is said to be \emph{input-to-state stable} (ISS) if there exist some $\Xi \in \mathcal{K}\mathcal{L}$ and $\phi \in \mathcal{K}_{\infty}$ functions \cite{khalil2002nonlinear} such that \eqref{eqISS}:
	\begin{equation}
		\Vert \bm{x}(t) \Vert \leqslant \Xi(\Vert \bm{x}(t=0) \Vert,t) + \phi(\Vert\bm{u}\Vert_\infty),
		\label{eqISS}
	\end{equation}
	is satisfied for all bounded inputs $\bm{u}$, all initial conditions $\bm{x}(t=0)$, and all $t\geqslant 0$. \\
\end{definition}

\subsection{ASV Dynamics} 
\label{subsecdynamics}

For an ASV moving in a horizontal plane, the kinematic and kinetic dynamics can be modelled as in \eqref{eqFossenModel} \cite{fossen2011handbook}:
\begin{equation}
	\begin{aligned}
		\left\{ 
		\begin{matrix}
			\begin{aligned}
				\bm{\dot{\eta}} &= \bm{R}(\psi) \bm{\nu} \\
				\bm{M} \bm{\dot{\nu}} &= - \bm{C}(\bm{\nu})\bm{\nu} - \bm{D}(\bm{\nu})\bm{\nu} - \bm{g}(\bm{\nu},\bm{\eta}) + \bm{\tau_{w}} + \bm{\tau} , \\
			\end{aligned}
		\end{matrix}
		\right.
	\end{aligned}
	\label{eqFossenModel}
\end{equation}
where $\bm{\eta} = [x \quad y \quad \psi]^{T}$ is the position/heading vector in two dimensions expressed in the north-east-down earth-fixed inertial frame \{n\}, $\bm{\nu} = [u \quad v \quad r]^{T}$ is the surge-sway-yaw velocity vector expressed in the body-fixed frame \{b\}, $\bm{\tau} = [F_u \quad 0 \quad \tau_r]^{T}$ stands for the force/torque vector, which represents the control action\footnote{Without any loss of generality, it is assumed that the control action cannot affect the sway component.}, and $\bm{\tau_{w}} = \left[ F_{w,u} \quad F_{w,v} \quad \tau_{w,r} \right]^{T}$ refers to the environmental forces and torque due to wind, waves, and currents. The rotation matrix $\bm{R}(\psi)$ between the body-fixed frame \{b\} and the earth-fixed inertial frame \{n\} is given by \eqref{eqMatrixR}:
\begin{equation} 
	\bm{R}(\psi) = \left[
	\begin{matrix}
		\text{cos}(\psi) & -\text{sin}(\psi) & 0 \\
		\text{sin}(\psi) &  \text{cos}(\psi) & 0 \\
		0		  &	 0		   & 1 \\
	\end{matrix}
	\right] = 
	\left[
	\begin{matrix}
		\bm{R}_2(\psi) & \bm{0}_{2 \times 1} \\
		\bm{0}_{1 \times 2}		 & 1 \\
	\end{matrix}
	\right],
	\label{eqMatrixR}
\end{equation}
where $\bm{R}_2(\psi) \in \mathbb{R}^{2\times2}$ is a two-dimensional submatrix of the original rotation matrix $\bm{R}(\psi)$. Notice that, in \eqref{eqMatrixR} and in the remaining of the paper, $\bm{0}$ represents a matrix or vector of zeros with the appropriate  dimensions, while $\bm{I}$ represents an identity matrix with the appropriate dimensions. In \eqref{eqFossenModel}, $\bm{M} = \bm{M}^{T}$ is the inertia matrix, $\bm{C}(\bm{\nu})$ is the Coriolis and centrifugal matrix, $\bm{D}(\bm{\nu})$ is the damping matrix, and $\bm{g}(\bm{\eta},\bm{\nu})$ refers to unmodelled hydrodynamics and heave, pitch, and roll cross-coupling effects on the modelled degrees of freedom. The nomenclature used in \cite{skjetne2004nonlinear} for matrices $\bm{C}(\bm{\nu})$ and $\bm{D}(\bm{\nu})$ is applied. Regarding the inertia matrix, its structure is detailed in \eqref{eqMatrixM}, under the assumptions considered in \cite{fossen2011handbook}:
\begin{equation}
   \begin{aligned}
	\bm{M} = 
	\left[
	\begin{matrix}
		m - X_{\dot{u}} 		& 0 					& 0 	 				\\
		0 						& m - Y_{\dot{v}} 		& m x_g - Y_{\dot{r}} 	\\
		0 						& m x_g - N_{\dot{v}}   & I_z - N_{\dot{r}}  	\\
	\end{matrix}
	\right],
	\label{eqMatrixM}
	\end{aligned}
\end{equation}
where $m$ is the vessel total mass, $x_g$ is the distance from the centre of gravity of the vessel to the origin of the body-fixed frame, $I_z$ is the moment of inertia about the $Z_{b}$ axis, and $ X_{(\cdot)}$, $Y_{(\cdot)}$, and $N_{(\cdot)}$ are hydrodynamic parameters according to standard notation \cite{sname1950nomenclature}. Given that $\bm{M}$ is assumed to be symmetric, $Y_{\dot{r}}$ must match $N_{\dot{v}}$. 


As stated in \cite{liu2019state}, the so-called lumped generalised disturbances $\bm{\sigma} = \left[ \sigma_u \; \sigma_v \; \sigma_r \right]^T\in\mathbb{R}^3$ can be built as shown in \eqref{eqLiuSigma}:
\begin{equation}
	\begin{aligned}
		\bm{\sigma} := \bm{M}^{-1}\left[ -\bm{C}(\bm{\nu})\bm{\nu} - \bm{D}(\bm{\nu})\bm{\nu} - \bm{g}(\bm{\eta},\bm{\nu}) + \bm{\tau_w} \right], \\
	\end{aligned} 
	\label{eqLiuSigma}
\end{equation}
lumping together the hydrodynamic effects of the ASV, the internal unmodelled dynamics, coupling effects, and uncertainties, in addition to the external disturbances. According to \eqref{eqLiuSigma}, the ASV dynamics \eqref{eqFossenModel} can be rewritten as follows:
\begin{equation}
	\begin{aligned}
		\left\{ 
		\begin{matrix}
			\begin{aligned}
				\bm{\dot{\eta}} &= \bm{R}(\bm{\psi}) \bm{\nu}, \\
				\bm{\dot{\nu}} &= \bm{M}^{-1}\bm{\tau} + \bm{\sigma}. 
			\end{aligned}
		\end{matrix}
		\right.
	\end{aligned}
	\label{eqFossenModel_Liu}
\end{equation}

Furthermore, vector $\bm{\omega}$ is defined as the time derivative of the lumped 
42
 disturbances $\bm{\sigma}$, i.e. $\bm{\omega}\equiv \dot{\bm{\sigma}}$.

Concerning the observer design, the ASV dynamics are represented as in \eqref{eqFossenModel_Liu}, thus neglecting the known internal structure of the lumped generalised disturbance vector $\bm{\sigma}$ expressed in \eqref{eqLiuSigma}. This representation could be questionable from an engineering point of view, but notice that the accurate expression of $\bm{\sigma}$ indicated in \eqref{eqLiuSigma} turns out to depend on several hydrodynamic parameters, for instance, those included in the damping matrix $\bm{D}(\bm{\nu})$. The experimental identification of these hydrodynamic parameters is complex and the achieved accuracy may not be high enough to address accurate disturbance rejection \cite{skjetne2004nonlinear}. Moreover, the heave, pitch, and roll cross-coupling effects on the modelled degrees of freedom are usually difficult to identify, thus in general the model-based estimation of $\bm{\sigma}$ does not provide good results in practice \cite{liu2019state}. That is the reason why the observer design is based on the ASV dynamics \eqref{eqFossenModel_Liu}, instead of the standard representation \eqref{eqFossenModel}.




\begin{assumption} \label{assumptionBoundedSigma}
	The time derivative of the lumped disturbance vector $\bm{\sigma}$ satisfies $\Vert \bm{\omega} \Vert \leqslant \omega^{*}$, being $\omega^{*}$ a positive constant.
\end{assumption}

\begin{assumption} \label{assumptionUawLimits} 
	The yaw velocity $r$ is limited by some known minimum and maximum values $r_{\min}$ and $r_{\max}$.
\end{assumption}

\begin{remark}
	 Assumption \ref{assumptionBoundedSigma}, which has been used in similar works such as \cite{zhao2015extended,liu2019state}, is not particularly hard. Notice that, according to \eqref{eqLiuSigma}, $\dot{\sigma}$ has an affine link with the vessel acceleration $\dot{\nu}$, which must present some limits due to physical constraints of the propeller motors. Moreover, notice that Assumption \ref{assumptionBoundedSigma} does not imply the knowledge of the value of $\omega^{*}$. On the other hand, Assumption \ref{assumptionUawLimits} is not hard either. By means of a suitable identification or experimentation with the ASV, these bounds may be estimated. This assumption has been previously used in \cite{lindegaard2003acceleration,liu2019state}.
\end{remark}

The ASV is assumed to be equipped with a GPS and a gyrocompass, in such a way that only the current position and orientation in the earth-fixed inertial frame \{n\} can be measured, as in \eqref{eqOutput}:
\begin{equation} 
	\bm{y} = \bm{\eta} + \bm{n},
	\label{eqOutput}
\end{equation}
where $\bm{n} = \left[n_x \quad n_y \quad n_\psi \right]^T = \left[ \bm{n}_p \quad n_\psi \right]^T$ refers to the noise affecting the measured variables.

\begin{assumption} \label{assumptionBoundedNoise}
	No noise is assumed to affect the orientation measurement, thus $n_\psi=0$. The noise affecting the position measurement $\bm{n}_p$ satisfies $\Vert \bm{n}_p \Vert \leqslant n_p^{*}$, with constant $n_p^{*}>0$.
\end{assumption}

\begin{remark}
	Assumption \ref{assumptionBoundedNoise} is needed to derive some theoretical properties of the proposed observer structure. In particular, noiseless heading measurement is needed to ensure that the optimization stage of the observer design can be performed offline and only an updating law based on the accurate value of the heading $\psi$ must be applied online to compute the observer gain. The first part of Assumption \ref{assumptionBoundedNoise} has been used in similar works \cite{liu2019state}, for the same purpose. Concerning the position measurement, the assumption is softer than the one used in \cite{vasconcelos2010discrete,vasconcelos2011ins, peng2009adaptive}, since no particular statistical description of the exogenous signal is assumed to be known. Again, $n_p^{*}$ is not assumed to be known. Furthermore, this assumption is much softer than considering noiseless position measurement, as in \cite{liu2019state}. 
\end{remark}



\subsection{Proposed Observer Structure}
\label{subsecObserverStructure}

The ASV dynamics and measured outputs in \eqref{eqFossenModel_Liu} and \eqref{eqOutput} can be divided into positional and rotational parts, according to \eqref{eqPositionalDynamics} and \eqref{eqRotationalDynamics}. On the one hand, the positional non-linear dynamics are described by \eqref{eqPositionalDynamics}:
\begin{equation}
	\begin{aligned}
		\left\{ 
		\begin{matrix}
			&\bm{\dot{\chi}}_p = \bm{A}_p(\psi)\bm{\chi}_p + \bm{B}_p\bm{\tau} +  \bm{B}_{\omega,p}\bm{\omega}_p,\\
			&\bm{y}_p = \bm{\eta}_p + \bm{n}_p = \bm{C}_p\bm{\chi}_p + \bm{n}_p, \\
		\end{matrix}
		\right.
	\end{aligned} 
	\label{eqPositionalDynamics}
\end{equation}
where $\bm{\chi}_p \equiv \left[\bm{\eta}_p \quad \bm{\nu}_p \quad \bm{\sigma}_p \right]^T \in \mathbb{R}^6$ is the positional state vector, being $\bm{ \eta}_p \equiv \left[x \quad y \right]^T$, $\bm{\nu}_p \equiv \left[u \quad v \right]^T$, and $\bm{\sigma}_p \equiv \left[\sigma_u \quad \sigma_v \right]^T$, where $\sigma_u$ and $\sigma_v$ represent the lumped generalised disturbances affecting the surge and sway accelerations, respectively; $\bm{\omega}_p = \left[\omega_u \quad \omega_v \right]^T$ is the time derivative of the corresponding components of vector $\bm{\sigma}$. The positional dynamic matrices $\bm{A}_{p} \in \mathbb{R}^{6\times6}$, $\bm{B}_{p} \in \mathbb{R}^{6\times3}$, $\bm{B}_{\omega,p} \in \mathbb{R}^{6\times2}$, and $\bm{C}_{p} \in \mathbb{R}^{2\times6}$ are detailed in \eqref{eqPositionalStateSpaceMatrices}, where $\bm{M}^{-1}_p$ corresponds to the two first rows of $\bm{M}^{-1}$:
\begin{equation} 
	\begin{aligned} 
		\bm{A}_p(\psi) &= \left[
		\begin{matrix}
			\bm{0}_{2 \times 2}   	& \bm{R}_2(\psi) 	& \bm{0}_{2 \times 2}  \\
			\bm{0}_{2 \times 2} 		& \bm{0}_{2 \times 2} 		    & \bm{I}_{2 \times 2} \\
			\bm{0}_{2 \times 2} 		& \bm{0}_{2 \times 2} 		    & \bm{0}_{2 \times 2}  \\
		\end{matrix}
		\right], 
		\bm{B}_p = \left[
		\begin{matrix}
			\bm{0}_{2 \times 3}  \\
			\bm{M}^{-1}_p  \\
			\bm{0}_{2 \times 3} \\
		\end{matrix}
		\right],\\
		\bm{B}_{\omega,p} &= \left[
		\begin{matrix}
			\bm{0}_{2 \times 2} \\
			\bm{0}_{2 \times 2} \\
			\bm{I}_{2 \times 2} \\
		\end{matrix}
		\right], \quad
		\bm{C}_p = \left[
		\begin{matrix}
			\bm{I}_{2 \times 2} & \bm{0}_{2 \times 2} & \bm{0}_{2 \times 2} 
		\end{matrix}
		\right].
	\end{aligned}    
	\label{eqPositionalStateSpaceMatrices}
\end{equation}
On the other hand, the rotational linear dynamics are described by \eqref{eqRotationalDynamics}:
\begin{equation}
	\begin{aligned}
		\left\{ 
		\begin{matrix}
			&\bm{\dot{\chi}}_\psi = \bm{A}_{\psi}\bm{\chi}_\psi + \bm{B}_{\psi}\bm{\tau} +  \bm{B}_{\omega,\psi}\omega_\psi, \\
			&y_\psi = \psi = \bm{C}_{\psi}\bm{\chi}_\psi, \\
		\end{matrix}
		\right.
	\end{aligned} 
	\label{eqRotationalDynamics}
\end{equation}
where $\bm{\chi}_\psi = \left[\psi \quad r \quad \sigma_r \right]^T \in \mathbb{R}^3$ is the rotational state vector, $\omega_\psi$ is the time derivative of the lumped disturbance component $\sigma_r$. The rotational dynamic matrices $\bm{A}_{\psi} \in \mathbb{R}^{3\times3}$, $\bm{B}_{\psi} \in \mathbb{R}^{3\times3}$, $\bm{B}_{\omega,\psi} \in \mathbb{R}^3$, and $\bm{C}_{\psi} \in \mathbb{R}^{1\times3}$ are given in \eqref{eqRotationalStateSpaceMatrices}, where $\bm{M}^{-1}_\psi$ is the third row of $\bm{M}^{-1}$:
\begin{equation} 
	\begin{aligned} 
		\bm{A}_\psi &= \left[
		\begin{matrix}
			0   	& 1 			& 0  \\
			0 		& 0 		    & 1  \\
			0 		& 0 		    & 0  \\
		\end{matrix}
		\right], 
		\bm{B}_\psi = \left[
		\begin{matrix}
			\bm{0}_{1 \times 3}  \\
			\bm{M}^{-1}_\psi  \\
			\bm{0}_{1 \times 3} \\
		\end{matrix}
		\right], \\
		\bm{B}_{\omega,\psi} &= \left[
		\begin{matrix}
			0 \\
			0 \\
			1 \\
		\end{matrix}
		\right], \quad 
		\bm{C}_\psi = \left[
		\begin{matrix}
			1  &0  &0 
		\end{matrix}
		\right].
	\end{aligned}    
	\label{eqRotationalStateSpaceMatrices}
\end{equation}

The proposed decoupling reveals a cascade structure, since the rotational dynamics are independent, and the positional dynamics are coupled with the rotational ones through the rotation matrix $\bm{R}_2(\psi)$. This fact is exploited in the definition of the observer structure.

Given the rotational dynamics \eqref{eqRotationalDynamics}, the linear rotational observer shown in \eqref{eqRotationalObserver} is proposed: 
\begin{equation}
	\begin{aligned}
		\dot{\hat{\bm{\chi}}}_\psi = \bm{A}_{\psi}\hat{\bm{\chi}}_\psi + \bm{B}_{\psi} \bm{\tau} +  \bm{L}_{\psi}(y_\psi - \bm{C}_{\psi}\hat{\bm{\chi}}_\psi), 
	\end{aligned} 
	\label{eqRotationalObserver}
\end{equation}
where $\bm{L}_{\psi}$ is a constant observer gain to be designed.

Notice that the positional dynamics \eqref{eqPositionalDynamics} are coupled with the orientation of the ASV through the rotation submatrix $\bm{R}_2(\psi)$. This non-linear coupling prevents from presenting a completely linear observer, as for the rotational dynamics. Therefore, the non-linear positional observer shown in \eqref{eqPositionalObserver} is proposed: 
\begin{equation}
	\begin{aligned}
		\dot{\hat{\bm{\chi}}}_p = \bm{A}_p(\psi)\hat{\bm{\chi}}_p + \bm{B}_{p} \bm{\tau} + 
		\bm{L}_{p}(\psi)(\bm{y}_p - \bm{C}_p\hat{\bm{\chi}}_p), \\
	\end{aligned} 
	\label{eqPositionalObserver}
\end{equation}
where the time-varying observer gain $\bm{L}_p(\psi)$ must still be defined.

The rotational estimation error is defined as $\bm{e}_{\psi}\equiv \bm{\chi}_\psi-\hat{\bm{\chi}}_\psi$, and the positional estimation error as $\bm{e}_p\equiv \bm{\chi}_p-\hat{\bm{\chi}}_p$.


\begin{problem} \label{Problem_2}
	Consider an ASV with the dynamics indicated in \eqref{eqPositionalDynamics} and \eqref{eqRotationalDynamics} and equipped with sensors measuring the variables indicated in \eqref{eqOutput}. Assume that the  robust non-linear observer \eqref{eqRotationalObserver}-\eqref{eqPositionalObserver} is being implemented on the ASV to estimate both the rotational and positional state variables. Then, the objective is to design the constant observer gain $\bm{L}_{\psi}$ and the time-varying observer gain $\bm{L}_p(\psi)$, in such a way that:
	\begin{description}
		\item[(i)]  The dynamics of the rotational estimation error $\bm{e}_\psi$ are ISS with respect to $\omega_\psi$.
		\item[(ii)] The dynamics of the positional estimation error $\bm{e}_p$ are ISS with respect to $\bm{\omega}_p$ and $\bm{n}_p$.
	\end{description}
	In addition, among all feasible $\bm{L}_{\psi}$ and $\bm{L}_p(\psi)$, the gains will be chosen to minimize the asymptotic estimation error bounds $\|\bm{e}_\psi\|$ and $\|\bm{e}_p\|$, this is, the region in which the estimation errors will be contained for sure in steady state.
\end{problem} 

An additional objective is to present an estimator structure that can be tuned with meaningful scalars, whose influence can be easily understood by the practitioner. Notice that, in the presence of unknown internal/external disturbances (Assumption \ref{assumptionBoundedSigma}), as well as bounded measurement noise (Assumption \ref{assumptionBoundedNoise}), only input-to-state stability can be achieved. Asymptotic stability would be only achieved in the case that neither noise nor disturbances affect the vessel, but this is unrealistic in actual ASV applications.


\section{Observer Design and Analysis} 
\label{secObserverDesign}


\subsection{Rotational Observer Design and Analysis} 
\label{subsecRotationalObserver}

The constant observer gain $\bm{L}_\psi \in \mathbb{R}^{3}$ in \eqref{eqRotationalObserver} is proposed to be computed as in \eqref{eqRotationalObserverGain}:
\begin{equation}
	\begin{aligned}
		\bm{L}_\psi = - \bm{P}_\psi^{-1} \bm{W}_\psi,
	\end{aligned} 
	\label{eqRotationalObserverGain}
\end{equation}
where the matrices $\bm{P}_\psi \in \mathbb{R}^{3 \times 3}$ and $\bm{W}_\psi \in \mathbb{R}^{3}$ are part of the solution of the optimization problem presented in \eqref{eqRotationalOptimizationProblem}, subject to some LMIs: 
\begin{subequations}
	\begin{align}
		&\min_{\bm{P}_\psi,\bm{W}_\psi,\beta_\psi}        \;\; \beta_\psi\label{eqRotationalOptimizationProblemObjFunction} \\
		&\text{s.t.} \quad \bm{P}_\psi > 0,\label{eqRotationalOptimizationProblemPositiveDefiniteLMI}\\
		&   \qquad \bm{A}^{T}_{\psi}\bm{P}_\psi + \bm{P}_\psi \bm{A}_\psi
		+ \bm{C}_\psi^{T}\bm{W}_\psi^{T} + \bm{W}_\psi\bm{C}_\psi + \bm{I} \leqslant 0,\label{eqRotationalOptimizationProblemStabilityLMI}
		\\
		&      \qquad \left[
		\begin{matrix}
			-\beta_\psi\bm{I}  	& \bm{P}_\psi \bm{B}_{\omega,\psi}  \\
			*	  & -\beta_\psi \\
		\end{matrix}
		\right]\leqslant 0, \label{eqRotationalOptimizationProblemDisturbMinimizationLMI}
	\end{align}
	\label{eqRotationalOptimizationProblem}
\end{subequations}
where $\beta_{\psi}\in\mathbb{R}$ is a decision variable. Notice that an asterisk has been used to mark the transposed element of the LMI \eqref{eqRotationalOptimizationProblemDisturbMinimizationLMI}.
\begin{theorem}
 \label{theoremRotationalObserverISS} 
	Problem \ref{Problem_2} (i) is solved provided that the constant observer gain $\bm{L}_{\psi}$ is computed by solving the optimization problem \eqref{eqRotationalOptimizationProblem} and then applying \eqref{eqRotationalObserverGain}.
\end{theorem}

\begin{proof} \label{proofRotationalObserverISS} 
	The dynamics of the rotational estimation error can be computed as in \eqref{eqRotationalObserverErrorDynamics}:
	\begin{equation}
		\begin{aligned}
			\dot{\bm{e}}_\psi = \dot{\bm{\chi}}_\psi - \dot{\hat{\bm{\chi}}}_\psi 
			=&\bm{A}_{\psi}\bm{\chi}_\psi + \bm{B}_{\psi}\bm{\tau} + \bm{B}_{\omega,\psi}\omega_\psi \\
			&-(\bm{A}_{\psi} - \bm{L}_\psi\bm{C}_\psi)\hat{\bm{\chi}}_\psi - \bm{B}_\psi \bm{\tau} - \bm{L}_\psi\bm{C}_\psi\bm{\chi}_\psi \\
			=&(\bm{A}_{\psi} - \bm{L}_\psi \bm{C}_\psi)\bm{e}_\psi + \bm{B}_{\omega,\psi}\omega_\psi .
		\end{aligned} 
		\label{eqRotationalObserverErrorDynamics}
	\end{equation}
	To analyse the stability of the rotational estimation error dynamics indicated in \eqref{eqRotationalObserverErrorDynamics}, a Lyapunov function candidate is proposed as indicated in \eqref{eqRotationalLyapunovFunctionCandidate}, where $\bm{P}_\psi \in \mathbb{R}^{3\times3}$ is a positive definite matrix according to \eqref{eqRotationalOptimizationProblemPositiveDefiniteLMI}.
	\begin{equation}
		\begin{aligned}
			V(\bm{e}_\psi) = \dfrac{1}{2}\bm{e}_\psi^{T}\bm{P}_\psi\bm{e}_\psi.
		\end{aligned} 
		\label{eqRotationalLyapunovFunctionCandidate}
	\end{equation}
	The time derivative of the Lyapunov function candidate yields:
	\begin{equation}
		\begin{aligned}
			\dot{V}(\bm{e}_\psi)=&\dfrac{1}{2}\bm{e}_\psi^{T}\left[(\bm{A}_{\psi} - \bm{L}_\psi\bm{C}_\psi)^{T}\bm{P}_\psi + \bm{P}_\psi(\bm{A}_{\psi} - \bm{L}_\psi\bm{C}_\psi)\right]\bm{e}_\psi \\ &+\bm{e}_\psi^{T}\bm{P}_\psi\bm{B}_{\omega,\psi}\omega_\psi.
		\end{aligned} 
		\label{eqRotationalStabilityAnalysis}
	\end{equation}
	By defining $\bm{W}_\psi \equiv - \bm{P}_\psi\bm{L}_\psi$, the time derivative of the Lyapunov function candidate indicated in  \eqref{eqRotationalStabilityAnalysis} can be expressed as follows:
	\begin{equation}
		\begin{aligned}
			\dot{V}(\bm{e}_\psi)
			= &\dfrac{1}{2}\bm{e}_\psi^{T}\left[\bm{A}^{T}_{\psi}\bm{P}_\psi + \bm{P}_\psi \bm{A}_\psi + \bm{C}_\psi^{T}\bm{W}_\psi^{T} + \bm{W}_\psi\bm{C}_\psi\right]\bm{e}_\psi \nonumber \\ 
			&+\bm{e}_\psi^{T}\bm{P}_\psi\bm{B}_{\omega,\psi}\omega_\psi .
		\end{aligned} 
		\label{eqStabilityAnalysisVariableChange}
	\end{equation}
	According to \cite{lindegaard2003acceleration} and standard norm properties, if \eqref{eqRotationalOptimizationProblemStabilityLMI} is satisfied, it implies that:
    \begin{equation}
		\begin{aligned}
			\dot{V}(\bm{e}_\psi) 
			&\leqslant -\dfrac{1}{2} \Vert\bm{e}_\psi \Vert^2 + \Vert\bm{e}_\psi\Vert \Vert \bm{P}_\psi \bm{B}_{\psi}\Vert \Vert\bm{\omega}_{\psi}\Vert.
		\end{aligned} 
		\label{eqRotationalLyapunovInequality}
	\end{equation}
	The inequality \eqref{eqRotationalLyapunovInequality} implies that:
	\begin{equation}
		\begin{aligned}
			\dot{V}(\bm{e}_\psi) \leqslant -\dfrac{1}{2} (1 - \theta_\psi)\Vert\bm{e}_\psi\Vert^2, 
		\end{aligned} 
		\label{eqRotationalLyapunovInequality2}
	\end{equation}
	where $0<\theta_\psi<1$, whenever:
	\begin{equation}
		\begin{aligned}
			\Vert\bm{e}_\psi\Vert  \geqslant \dfrac{2 \Vert \bm{P}_\psi\bm{B}_{\psi} \Vert \Vert\bm{\omega}_{\psi}\Vert}{\theta_\psi} .
		\end{aligned} 
		\label{eqRotationalErrorBound}
	\end{equation}
	Therefore, according to Assumption \ref{assumptionBoundedSigma}, the rotational estimation error dynamics \eqref{eqRotationalObserverErrorDynamics} are ISS with respect to $\bm{\omega}_{\psi}$ and:
	\begin{equation}
		\begin{aligned}
			\Vert\bm{e}_\psi\Vert (t) \leqslant
			\text{max} \{ \Xi_\psi(\Vert\bm{e}_\psi(0)\Vert,t),\;\phi_\psi(\Vert\bm{\omega}_{\psi}\Vert) \},
		\end{aligned} 
		\label{eqRotationalErrorBound2}
	\end{equation}
	where $\Xi_\psi$ is a $\mathcal{K}\mathcal{L}$ function \cite{khalil2002nonlinear} and $\phi_\psi(\Vert\bm{\omega}_{\psi}\Vert)$ is a $\mathcal{K}_\infty$ function defined as:
	\begin{equation}
		\begin{aligned}
			\phi_\psi(\Vert\bm{\omega}_{\psi}\Vert) \equiv \sqrt{\dfrac{\lambda_{\max}(\bm{P}_\psi)}{\lambda_{\min}(\bm{P}_\psi)}}\dfrac{2\Vert\bm{P}_\psi\bm{B}_{\psi}\Vert}{\theta_\psi} \Vert\bm{\omega}_{\psi}\Vert.
		\end{aligned} 
		\label{eqRotationalDisturbanceFunction}
	\end{equation}
	
	Notice that $\bm{\omega}_{\psi}$ is bounded due to Assumption \ref{assumptionBoundedSigma}. Hence, the rotational error dynamics are ISS with respect to $\omega_{\psi}$. 
	
	Moreover, the asymptotic or final estimation error bound $\Vert\bm{e}_\psi\Vert$ can be reduced by minimizing $\Vert\bm{P}_\psi\bm{B}_{\psi}\Vert$, or, equivalently, minimizing the decision variable $\beta_\psi$ such as $\Vert\bm{P}_\psi\bm{B}_{\psi}\Vert \leqslant \beta_\psi$. This can be done by solving the LMI-based optimization problem shown in \eqref{eqRotationalOptimizationProblem}, where the Schur complement \cite{zhang2006schur} has been applied to obtain the constraint indicated in \eqref{eqRotationalOptimizationProblemDisturbMinimizationLMI} related to the condition $\Vert\bm{P}_\psi\bm{B}_{\psi}\Vert \leqslant \beta_\psi$. Once the optimization problem is solved for the decision variables $\bm{P}_\psi$, $\bm{W}_\psi$, and $\beta_\psi$, the observer gain $\bm{L}_\psi$ can be computed by applying \eqref{eqRotationalObserverGain}. 
\end{proof} 

\begin{remark} \label{remarkLMIComputation}
	The LMI optimization problem \eqref{eqRotationalOptimizationProblem} can easily be solved using any LMI software package, such as SeDuMi \cite{SEDUMI} and YALMIP \cite{YALMIP}. Moreover, both the optimization problem \eqref{eqRotationalOptimizationProblem} and the computation of the rotational observer gain \eqref{eqRotationalObserverGain} are performed only once and offline, in such a way that only the observer dynamics \eqref{eqRotationalObserver} must be implemented online, thus not compromising the computational feasibility of the observer, even on limited-power single-board computers.
\end{remark} 

\begin{remark} \label{remarkPolePlacement}
	A given velocity of convergence of the rotational observer can be ensured by placing its continuous-time poles at specific locations in the complex plane. If the real part of all poles is intended to be located on the left of $-\delta_{\psi,1}$, $\Re(\varphi_\psi)\leqslant -\delta_{\psi,1}$, where $\delta_{\psi,1} \in \mathbb{R} \geqslant 0$ and $\varphi_\psi \in \mathbb{C} $ are the rotational error dynamics poles, the LMI \eqref{eqRotationalOptimizationProblemPolePlacementLMI} can be included in the optimization problem \eqref{eqRotationalOptimizationProblem} as an additional constraint \cite{chilali1999robust}:
	\begin{equation}
		\begin{aligned}
			\begin{matrix}
				\bm{A}^{T}_{\psi}\bm{P}_\psi + \bm{P}_\psi\bm{A}_\psi + \bm{C}_\psi^{T}\bm{W}_\psi^{T} + \bm{W}_\psi\bm{C}_\psi + 2\delta_{\psi,1}\bm{P}_\psi \leqslant 0.
			\end{matrix}
		\end{aligned} 
		\label{eqRotationalOptimizationProblemPolePlacementLMI}
	\end{equation}
	
\end{remark} 

The value of $\delta_{\psi,1}$ is the only tuning parameter of the rotational observer, related to the minimum required velocity of convergence. Indeed, the inverse of $\delta_{\psi,1}$ is related to the time constant of the convergence of the rotational observer to the actual values of the estimated variables.

Therefore, solving the optimization problem \eqref{eqRotationalOptimizationProblem} with objective function \eqref{eqRotationalOptimizationProblemObjFunction} and constraints \eqref{eqRotationalOptimizationProblemPositiveDefiniteLMI}-\eqref{eqRotationalOptimizationProblemDisturbMinimizationLMI} and \eqref{eqRotationalOptimizationProblemPolePlacementLMI} gives rise to an ISS rotational observer with respect to the disturbance $\omega_\psi$, with minimization of the estimation error bound and velocity of convergence defined by the tuning parameter $\delta_{\psi,1}$. 

\begin{remark} \label{remarkFeasibility}
	Since the pair ($\bm{A}_{\psi}$, $\bm{C}_{\psi}$) is observable, the optimization problem \eqref{eqRotationalOptimizationProblem} is feasible. Note that $\beta_{\psi}$ can be made as big as needed to satisfy \eqref{eqRotationalOptimizationProblemDisturbMinimizationLMI}. The feasibility is not compromised even if the additional constraint \eqref{eqRotationalOptimizationProblemPolePlacementLMI} is considered, since, for a linear observable system, the poles of the error dynamics can be located anywhere. Furthermore, the faster the poles are chosen, the bigger the final error bound will be.
\end{remark}

\subsection{Positional Observer Design and Analysis} 
\label{subsecPositionalObserver}
The time-varying observer gain $\bm{L}_p(\psi) \in \mathbb{R}^{6 \times 2}$ is proposed to be continuously updated as in \eqref{eqPositionalObserverGain}:
\begin{equation}
	\begin{aligned}
		\bm{L}_p(\psi) = - \bm{T}_p^{-1}\bm{P}_p^{-1} \bm{W}_p \bm{R}_{2}^T(\psi),
	\end{aligned} 
	\label{eqPositionalObserverGain}
\end{equation}
being $\bm{T}_p \in \mathbb{R}^{6\times6}$ a block-diagonal coordinate transformation matrix, defined as indicated in \eqref{eqMatrixT}:
\begin{equation}
	\begin{aligned}
		\bm{T}_p = \left[
		\begin{matrix}
			\bm{R}_2^{T}(\psi)   & \bm{0}_{2 \times 2} 		& \bm{0}_{2 \times 2} \\
			\bm{0}_{2 \times 2} 		                 & \bm{I}_{2 \times 2} 	& \bm{0}_{2 \times 2} \\
			\bm{0}_{2 \times 2} 		                 & \bm{0}_{2 \times 2} 		& \bm{I}_{2 \times 2}\\
		\end{matrix}
		\right].
	\end{aligned}
	\label{eqMatrixT}
\end{equation}

Constant matrices $\bm{P}_p \in \mathbb{R}^{6 \times 6}$ and $\bm{W}_p \in \mathbb{R}^{6 \times 2}$ are part of the solution of the LMI optimization problem \eqref{eqPositionalOptimizationProblem}:
\begin{subequations}
	\begin{align}
		&\min_{\bm{P}_p,\bm{W}_p,\beta_p}    \;\;    \beta_p\label{eqPositionalOptimizationProblemObjFunction} \\
		&\text{s.t.}  \quad  \bm{P}_p > 0,\label{eqPositionalOptimizationProblemPositiveDefiniteLMI}
		\\
		&\qquad \begin{aligned}
			&\bm{A}^T_{0,p}\bm{P}_p+\bm{P}_p\bm{A}_{0,p} +\bm{C}_p^{T}\bm{W}_p^T +\bm{W}_p\bm{C}_p  \\
			&+r_{\min}(\bm{S}^{T}_{T_p}\bm{P}_p + \bm{P}_p\bm{S}_{T_p}) + \bm{I} \leqslant 0, \\ 
			  \end{aligned} \label{eqPositionalOptimizationProblemStabilityLMI1} \\
		&\qquad \begin{aligned} 
			&\bm{A}^T_{0,p}\bm{P}_p+\bm{P}_p\bm{A}_{0,p} +\bm{C}_p^{T}\bm{W}_p^T +\bm{W}_p\bm{C}_p \\
			&+ r_{\max}(\bm{S}^{T}_{T_p}\bm{P}_p + \bm{P}_p\bm{S}_{T_p}) + \bm{I} \leqslant 0, \\
			  \end{aligned} \label{eqPositionalOptimizationProblemStabilityLMI2} \\
		&\quad      
		\begingroup 
		\setlength\arraycolsep{3.5pt}
		\begin{bmatrix}
			-\beta_p\bm{I}  	& \bm{P}_p\bm{T}_p\bm{B}_{\omega,p}k_{\omega,p}  &\bm{W}_p k_{n,p}  \\
			*	  & -\beta_p\bm{I}  & \bm{0} \\
			* & * & -\beta_p\bm{I}  \\
		\end{bmatrix}
		\endgroup
		\leqslant 0,\label{eqPositionalOptimizationProblemTradeoffLMI}
	\end{align}
	\label{eqPositionalOptimizationProblem}
\end{subequations}
where $\beta_{p}\in\mathbb{R}$ is a decision variable, $k_{\omega,p} \in \mathbb{R}\geqslant 0$ the weighting parameter for robustness against disturbances and $k_{n,p}\in\mathbb{R}\geqslant 0$ the weighting parameter for noise rejection, and
\begin{equation} 
	\begin{aligned}
		\bm{A}_{0,p} &\equiv 
		\left[    
		\begin{matrix}
			\bm{0}_{2 \times 2}   	& \bm{I}_{2 \times 2} 		& \bm{0}_{2 \times 2}  \\
			\bm{0}_{2 \times 2}		& \bm{0}_{2 \times 2}		& \bm{I}_{2 \times 2} \\
			\bm{0}_{2 \times 2} 	& \bm{0}_{2 \times 2}	    & \bm{0}_{2 \times 2}  \\
		\end{matrix}
		\right], \\
		\bm{S}_{T_p} &\equiv \left[
		\begin{matrix}
			\bm{S}^{T}_p   & \bm{0}_{2 \times 2}  	& \bm{0}_{2 \times 2}  \\
			\bm{0}_{2 \times 2} 		       & \bm{0}_{2 \times 2} 	& \bm{0}_{2 \times 2}  \\
			\bm{0}_{2 \times 2} 		       & \bm{0}_{2 \times 2}   & \bm{0}_{2 \times 2} \\
		\end{matrix}
		\right], \; \bm{S}_p =\left[
		\begin{matrix}
			0        & -1 \\
			1 		 &  0 \\
		\end{matrix}
		\right].
	\end{aligned}   
	\label{eqMatrixAp0}
\end{equation}
Notice that $\bm{T}_p\bm{B}_{\omega,p} = \bm{B}_{\omega,p}$, thus the LMI \eqref{eqPositionalOptimizationProblemTradeoffLMI} is well posed.

Please also notice that, in order to compute the observer gain \eqref{eqPositionalObserverGain}, the heading $\psi$ must be accurately known, which is true according to Assumption \ref{assumptionBoundedNoise}. From this value, one can obtain the matrices $\bm{T}_p^{-1}$ and $\bm{R}_2(\psi)$, necessary to compute the observer gain $\bm{L}_p$ shown in \eqref{eqPositionalObserverGain}.


It is worth mentioning that the optimization problem \eqref{eqPositionalOptimizationProblem} must be only solved once and offline to get the constant matrices $\bm{P}_p$ and $\bm{W}_p$. Afterwards, the observer gain $\bm{L}_p(\psi)$ can be easily updated with \eqref{eqPositionalObserverGain}. Notice that only this updating law and the positional observer dynamics \eqref{eqPositionalObserver} must be implemented online.

\begin{theorem} \label{theoremPositionalObserverISS} 
	Problem \ref{Problem_2} (ii) is solved provided that the time-varying observer gain $\bm{L}_{p} (\psi)$ is computed by initially solving the optimization problem \eqref{eqPositionalOptimizationProblem} and then applying \eqref{eqPositionalObserverGain} in a continuous manner.
\end{theorem}

\begin{proof} \label{proofPositionalObserverISS} 
	The dynamics of the positional error can be computed as indicated in \eqref{eqObserverErrorDynamicsposition}:
	\begin{equation}
		\begin{aligned}
			\dot{\bm{e}}_p =&  \dot{\bm{\chi}}_p - \dot{\hat{\bm{\chi}}}_p = \\
			=&\bm{A}_{p}(\psi)\bm{\chi}_p + \bm{B}_p\bm{\tau} + \bm{B}_{\omega,p}\bm{\omega}_p - 
			(\bm{A}_{p}(\psi) - \bm{L}_p\bm{C}_p)\hat{\bm{\chi}}_p \\
			&-\bm{B}_p\bm{\tau} - \bm{L}_{p}\bm{C}_p\bm{\chi}_p - \bm{L}_{p} \bm{n}_p = \\
			=&(\bm{A}_{p}(\psi) - \bm{L}_{p}\bm{C}_p)\bm{e}_p + \bm{B}_{\omega,p}\bm{\omega}_p - \bm{L}_{p}\bm{n}_p.
		\end{aligned} 
		\label{eqObserverErrorDynamicsXposition}
	\end{equation}
	Because of the presence of $\psi$, the positional dynamics turn out to be non-linear. In order to analyse the stability properties of the estimation error, a coordinate transformation will be applied to the positional error. Then, using the transformation matrix previously defined in \eqref{eqMatrixT}, one can define:
	\begin{equation}
		\begin{aligned}
			&\bm{z}_p \equiv \bm{T}_p\bm{e}_p.
		\end{aligned} 
		\label{eqTransformation1}
	\end{equation}
	The time derivative of this transformed positional error is given in \eqref{eqTransformationDerivative}. Notice the presence of the time derivative of $\dot{\bm{T}}_p$, as this is not a constant transformation:
	\begin{equation}
		\begin{aligned}
			&\dot{\bm{z}}_p = \bm{T}_p \dot{\bm{e}}_p + \dot{\bm{T}}_p \bm{e}_p, 
		\end{aligned}
		\label{eqTransformationDerivative}
	\end{equation}
	where $\dot{\bm{T}}_p$ is given by
	\eqref{eqderivativetransformation}:
	\begin{equation}
		\begin{aligned}
			&\dot{\bm{T}}_p = r\bm{S}_{T_p}\bm{T}_p.
		\end{aligned} 
		\label{eqderivativetransformation}
	\end{equation}
	From \eqref{eqObserverErrorDynamicsXposition}-\eqref{eqTransformation1}, the evolution of the transformed positional error can be computed:
	\begin{equation}
		\begin{aligned}
			\dot{\bm{z}}_p
			=&\bm{T}_p(\bm{A}_{p}(\psi) - \bm{L}_{p}\bm{C}_p)\bm{T}_p^{-1}\bm{z}_p + r\bm{S}_{T_p}\bm{z}_p + \bm{T}_p\bm{B}_{\omega,p}\bm{\omega}_p  \\
			&-\bm{T}_p\bm{L}_{p}\bm{n}_p.
		\end{aligned} 
		\label{eqObserverErrorDynamicsposition}
	\end{equation}
	Notice that 
	\begin{equation} 
		\begin{aligned}
			\bm{T}_p \bm{A}_{p}(\psi)\bm{T}_p^{-1} 
			= \left[    
			\begin{matrix}
				\bm{0}_{2 \times 2}   	& \bm{I}_{2 \times 2}		& \bm{0}_{2 \times 2}  \\
				\bm{0}_{2 \times 2}		& \bm{0}_{2 \times 2}		& \bm{I}_{2 \times 2} \\
				\bm{0}_{2 \times 2} 	& \bm{0}_{2 \times 2}		& \bm{0}_{2 \times 2}  \\
			\end{matrix}
			\right] = \bm{A}_{0,p},
		\end{aligned}   
		\label{eqMatrixAp0_2}
	\end{equation}
	which was defined in \eqref{eqMatrixAp0}. In a similar way,
	\begin{equation}
		\begin{aligned}
			\bm{C}_p\bm{T}^{-1}_p &= \left[
			\begin{matrix}
				\bm{I}_{2 \times 2} & \bm{0}_{2 \times 2} & \bm{0}_{2 \times 2} 
			\end{matrix}
			\right]
			\left[
			\begin{matrix}
				\bm{R}_2(\psi)   & \bm{0}_{2 \times 2} 		 & \bm{0}_{2 \times 2} \\
				\bm{0}_{2 \times 2} 		     & \bm{I}_{2 \times 2} 	     & \bm{0}_{2 \times 2} \\
				\bm{0}_{2 \times 2} 		     & \bm{0}_{2 \times 2}        & \bm{I}_{2 \times 2}\\
			\end{matrix}
			\right] \\
			&=\left[
			\begin{matrix}
				\bm{R}_2(\psi)   & \bm{0}_{2 \times 2} 		 & \bm{0}_{2 \times 2} \\
			\end{matrix}
			\right] =\bm{R}_2(\psi) \bm{C}_p.
		\end{aligned} 
		\label{eqMatrxCR}
	\end{equation}	
	By defining $\bm{L}_{p_z} \equiv - \bm{T}_p \bm{L}_p \bm{R}_2(\psi)$, the dynamics of the transformed positional error can be rewritten as in \eqref{eqnewErrorDynamicsposition}:
	\begin{equation}
		\begin{aligned}
			\dot{\bm{z}}_p=&(\bm{A}_{0,p} + r\bm{S}_{T_p}+ \bm{L}_{p_z}\bm{C}_p)\bm{z}_p + \bm{T}_p\bm{B}_{\omega,p}\bm{\omega}_p \\
			&+\bm{L}_{p_z}\bm{R}_2^{T}(\psi)\bm{n}_p .
		\end{aligned} 
		\label{eqnewErrorDynamicsposition}
	\end{equation}
	In order to analyse the stability of the transformed positional estimation error dynamics \eqref{eqnewErrorDynamicsposition}, a Lyapunov function candidate is proposed in \eqref{eqPositionalLyapunovFunctionCandidate}, where $\bm{P}_p \in \mathbb{R}^{6\times6}$ is a positive definite matrix according to \eqref{eqPositionalOptimizationProblemPositiveDefiniteLMI}. Notice that, using the proposed coordinate transformation \eqref{eqTransformation1} and the definition of the positional observer gain \eqref{eqPositionalObserverGain}, the transformed error dynamics \eqref{eqnewErrorDynamicsposition} are linear, and thus a quadratic Lyapunov function can be proposed to analyse their stability.
	\begin{equation}
		\begin{aligned}
			V(\bm{z}_p) = \dfrac{1}{2}\bm{z}_p^{T}\bm{P}_p\bm{z}_p.
		\end{aligned} 
		\label{eqPositionalLyapunovFunctionCandidate}
	\end{equation}
	The time derivative of the Lyapunov function \eqref{eqPositionalLyapunovFunctionCandidate} yields:
	\begin{equation}
		\begin{aligned}
			\dot{V}(\bm{z}_p) =& 
			\dfrac{1}{2}\bm{z}_p^{T}\left(\bm{A}^{T}_{0,p}\bm{P}_p + \bm{P}_p\bm{A}_{0,p} + \bm{C}_p^{T}\bm{L}_{p_z}^{T}\bm{P}_p + \bm{P}_p\bm{L}_{p_z}\bm{C}_p\right)\\
			&+ \dfrac{1}{2}\bm{z}_p^{T}\left(r(\bm{S}^{T}_{T_p}\bm{P}_p + \bm{P}_p\bm{S}_{T_p})\right)\bm{z}_p +
			\bm{z}_p^{T}\bm{P}_p\bm{T}_p\bm{B}_{\omega,p}\bm{\omega}_p\\ &+\bm{z}_p^{T}\bm{P}_p\bm{L}_{p_z}\bm{R}_2^{T}(\psi)\bm{n}_p .
		\end{aligned} 
		\label{eqPositionalStabilityAnalysis}
	\end{equation}
	By defining $\bm{W}_p \equiv \bm{P}_p\bm{L}_{p_z}$, $\bm{\omega}_p \equiv k_{\omega,p} \bm{\omega}^{'}_p$, and $\bm{n}_p \equiv k_{n,p} \bm{n}^{'}_p$,   \eqref{eqPositionalStabilityAnalysis} can be expressed as follows:
			\begin{equation}
				\begin{aligned}
					&\dot{V}(\bm{z}_p) = \dfrac{1}{2}\bm{z}^{T}_p\left[\bm{A}^{T}_{0,p}\bm{P}_p + \bm{P}_p\bm{A}_{0,p} +\bm{C}_p^{T}\bm{W}^{T}_p + \bm{W}_p\bm{C}_p \right. \\
					&\left. +r(\bm{S}^{T}_{T_p}\bm{P}_p +\bm{P}_p\bm{S}_{T_p})\right]\bm{z}_p \\  
					&+ \bm{z}^{T}_p
					\left[
					\begin{matrix}	\bm{P}_p\bm{T}_p\bm{B}_{\omega,p}k_{\omega,p}  & \bm{W}_p k_{n,p} \\
					\end{matrix}
					\right] \cdot \left[
					\begin{matrix}
						\bm{I}_{2 \times 2}    & \bm{0}_{2 \times 2}  \\
						\bm{0}_{2 \times 2}    & \bm{R}^{T}_2(\psi)  \\
					\end{matrix}
					\right] 
					\left[
					\begin{matrix}
						\bm{\omega}^{'}_p \\
						\bm{n}^{'}_p \\
					\end{matrix}
					\right] = \\
					&= \dfrac{1}{2}\bm{z}^{T}_p\left[\bm{A}^{T}_{0,p}\bm{P}_p + \bm{P}_p\bm{A}_{0,p} +\bm{C}_p^{T}\bm{W}^{T}_p + \bm{W}_p\bm{C}_p \right. \\
					&\left. +r(\bm{S}^{T}_{T_p}\bm{P}_p +\bm{P}_p\bm{S}_{T_p})\right]\bm{z}_p \\
					&+ \bm{z}^{T}_p
					\begingroup 
					\setlength\arraycolsep{3.5pt}
					\begin{bmatrix}	\bm{P}_p\bm{T}_p\bm{B}_{\omega,p}k_{\omega,p}  & \bm{W}_p k_{n,p} \\
					\end{bmatrix}
					\endgroup
					\bm{G}_{p} \bm{\Gamma}^{'}_{p},
				\end{aligned} 
				\label{eqerrodynamicspositionWk}
			\end{equation}
	where $k_{\omega,p} \in \mathbb{R} \geqslant 0$ is the weighting parameter for robustness against disturbances and $k_{n,p} \in \mathbb{R} \geqslant 0$ is the weighting parameter for noise rejection, and:
	\begin{equation}
		\begin{aligned}
			&\bm{G}_{p} \equiv \left[
			\begin{matrix}
				\bm{I}_{2 \times 2}    & \bm{0}_{2 \times 2}  \\
				\bm{0}_{2 \times 2}    & \bm{R}^{T}_2(\psi) \\
			\end{matrix}
			\right] \in \mathbb{R}^{4 \times 4}, \\ 
			&\bm{\Gamma}^{'}_{p}\equiv\left[
			\begin{matrix}
				\bm{\omega}^{'}_p \\
				\bm{n}^{'}_p \\
			\end{matrix}
			\right] \in \mathbb{R}^{4}.
		\end{aligned} 
		\label{eqErrordynamicsPositionMG2}
	\end{equation}
	Once again according to \cite{lindegaard2003acceleration} and standard norm properties, if \eqref{eqPositionalOptimizationProblemStabilityLMI1} and \eqref{eqPositionalOptimizationProblemStabilityLMI2} are satisfied, it implies that:
	\begin{equation}
		\begin{aligned}
			\dot{V}(\bm{z}_p) &\leqslant -\dfrac{1}{2} \Vert\bm{z}_p \Vert^2 + \Vert\bm{z}_p\Vert
			\cdot \Vert 
			\begingroup 
			\setlength\arraycolsep{3.5pt}
			\begin{bmatrix}
			    \bm{P}_p\bm{T}_p\bm{B}_{\omega,p}k_{\omega,p}  & \bm{W}_p k_{n,p} 
			\end{bmatrix}
			\endgroup
			\Vert \\
			& \cdot \Vert\bm{G}_{p} \Vert 
			\Vert\bm{\Gamma}^{'}_{p}\Vert  
			\leqslant -\dfrac{1}{2} \Vert\bm{z}_p \Vert^2 + \Vert\bm{z}_p\Vert \Vert\bm{\Upsilon}_{p}\Vert \Vert\bm{\Gamma}^{'}_{p}\Vert,
		\end{aligned} 
		\label{eqPositionalLyapunovInequality}
	\end{equation}
	where 
	\begin{equation}
		\bm{\Upsilon}_{p} \equiv 
		\begingroup 
		\setlength\arraycolsep{2.5pt} 
		\begin{bmatrix} 
			\bm{P}_p\bm{T}_p\bm{B}_{\omega,p}k_{\omega,p} \quad \bm{W}_p k_{n,p}.
		\end{bmatrix} 
	\endgroup 
	\label{eqPositionalLyapunovGammaDefinition}
	\end{equation} 
	Notice that $\Vert\bm{G}_{p} \Vert \leq 1$, given the definition of the rotation submatrix $\bm{R}_2(\psi)$. The inequality \eqref{eqPositionalLyapunovInequality} implies that:
	\begin{equation}
		\begin{aligned}
			\dot{V}(\bm{z}_p) \leqslant -\dfrac{1}{2} (1 - \theta_z)\Vert\bm{z}_p\Vert^2, 
		\end{aligned} 
		\label{eqPositionalLyapunovInequality2}
	\end{equation}
	where $0<\theta_z<1$, whenever:
	\begin{equation}
		\begin{aligned}
			\Vert\bm{z}_p\Vert  \geqslant \dfrac{2\Vert\bm{\Upsilon}_{p}\Vert \Vert\bm{\Gamma}^{'}_{p}\Vert}{\theta_z}.
		\end{aligned} 
		\label{eqPositionalErrorBound}
	\end{equation}
	Therefore, according to Assumption \ref{assumptionBoundedSigma}, the estimation error dynamics \eqref{eqnewErrorDynamicsposition} are ISS with respect to $\bm{\Gamma}^{'}_{p}$:
	\begin{equation}
		\begin{aligned}
			\Vert\bm{z}_p\Vert (t) \leqslant
			\text{max} \{\Xi_p(\Vert\bm{z}_p(0)\Vert,t),\;\phi(\Vert\bm{\Gamma}^{'}_{p}\Vert)\},
		\end{aligned} 
		\label{eqPositionalErrorBound2}
	\end{equation}
	where $\Xi_p$ is a $\mathcal{K}\mathcal{L}$ function \cite{khalil2002nonlinear} and $\phi_p(\Vert\bm{\Gamma}^{'}_p\Vert)$ is a $\mathcal{K}_\infty$ function defined as
	\begin{equation}
		\begin{aligned}
			\phi_p(\Vert\bm{\Gamma}^{'}_{p}\Vert) \equiv \sqrt{\dfrac{\lambda_{\max}(\bm{P}_p)}{\lambda_{\min}(\bm{P}_p)}}\dfrac{2\Vert\bm{\Upsilon}_{p}\Vert \Vert\bm{\Gamma}^{'}_{p}\Vert}{\theta_z}.
		\end{aligned} 
		\label{eqPositionalDisturbanceNoiseFunction}
	\end{equation}
	Notice that $\bm{\Gamma}^{'}_{p}$ is bounded due to the constraints imposed to $\bm{\omega}_{p}$ and $\bm{n}_p$ in Assumptions \ref{assumptionBoundedSigma} and \ref{assumptionBoundedNoise}, respectively. Hence, the transformed positional error dynamics are ISS with respect to $\bm{\Gamma}^{'}_{p}$, and in turn to $\bm{\Gamma}_p  \equiv \left[ \bm{\omega}_p \quad \bm{n}_p \right]^{T}$, since the difference between both vectors is only a scaling factor, which is used to allow the user to select the trade-off between robustness against uncertainties and noise rejection. Moreover, from the definition of the transformed error \eqref{eqTransformation1}, the dynamics of the original positional error $\bm{e}_p$ are also ISS with respect to $\bm{\Gamma}_p$.
	
	Finally, the proposed design must be proved to minimize the final bound of $\Vert \bm{e}_p \Vert$. It holds that $\Vert \bm{e}_p \Vert = \Vert \bm{T}_{p}^{-1} \bm{z}_{p} \Vert \leq \Vert \bm{T}_{p}^{-1}\Vert \Vert \bm{z}_{p} \Vert\leq \Vert \bm{z}_{p} \Vert$. Then, the minimization of $\Vert \bm{z}_{p} \Vert$ is sought. 
	
	The final estimation error bound $\Vert\bm{z}_p\Vert$ can be reduced by minimizing $\Vert\bm{\Upsilon}_{p}\Vert$ for given constants $k_{\omega,p}$, $k_{n,p}$, or equivalently minimizing $\beta_p$ such as $\Vert\bm{\Upsilon}_{p}\Vert \leqslant \beta_p$. It can be done by solving the optimization \eqref{eqPositionalOptimizationProblem}, where the Schur complement \cite{zhang2006schur} has been applied to obtain the constraint \eqref{eqPositionalOptimizationProblemTradeoffLMI}, related to the condition $\Vert\bm{\Upsilon}_{p}\Vert \leqslant \beta_p$. Once \eqref{eqPositionalOptimizationProblem} is solved for the decision variables $\bm{P}_p$, $\bm{W}_p$, and $\beta_p$, the time-varying observer gain is computed online by applying \eqref{eqPositionalObserverGain}.  
\end{proof}

The LMI-based optimization problem \eqref{eqPositionalOptimizationProblem} can  be solved using any LMI software package, such as those mentioned in Remark \ref{remarkLMIComputation}.

\begin{remark} \label{remarkPositionalPolePlacement}
	Once again, in the case of the positional observer a given velocity of convergence can be ensured by placing its continuous-time poles at specific locations in the complex plane. If the real part of all poles is intended to be located on the left of $-\delta_{p,1}$, $\Re(\varphi_p)\leqslant -\delta_{p,1}$, where $\delta_{p,1} \in \mathbb{R} \geqslant 0$ and $\varphi_p \in \mathbb{C} $ are the positional error dynamics poles, the LMIs indicated in \eqref{eqPositionalOptimizationProblemPolePlacementLMI} can be included in the optimization problem \eqref{eqPositionalOptimizationProblem} as additional constraints \cite{chilali1999robust}:
	\begin{equation}
		\begin{aligned}
			&\bm{A}^T_{0,p}\bm{P}_p+\bm{P}_p\bm{A}_{0,p} +\bm{C}_p^{T}\bm{W}_p^T + \bm{W}_p\bm{C}_p \\ &+r_{\min}(\bm{S}^{T}_{T_p}\bm{P}_p + \bm{P}_p\bm{S}_{T_p}) + 2\delta_{p,1}\bm{P}_p \leqslant 0, \\
			&\bm{A}^T_{0,p}\bm{P}_p+\bm{P}_p\bm{A}_{0,p}+\bm{C}_p^T\bm{W}_p^T + \bm{W}_p\bm{C}_p \\ &+r_{\max}(\bm{S}^T_{T_p}\bm{P}_p + \bm{P}_p\bm{S}_{T_p}) + 2\delta_{p,1}\bm{P}_p \leqslant 0. \\
		\end{aligned}
		\label{eqPositionalOptimizationProblemPolePlacementLMI}
	\end{equation}
\end{remark} 

\begin{figure*}[!ht]
	\centering
	\includegraphics[width=16.0cm,trim = 40 55 0 40,clip] {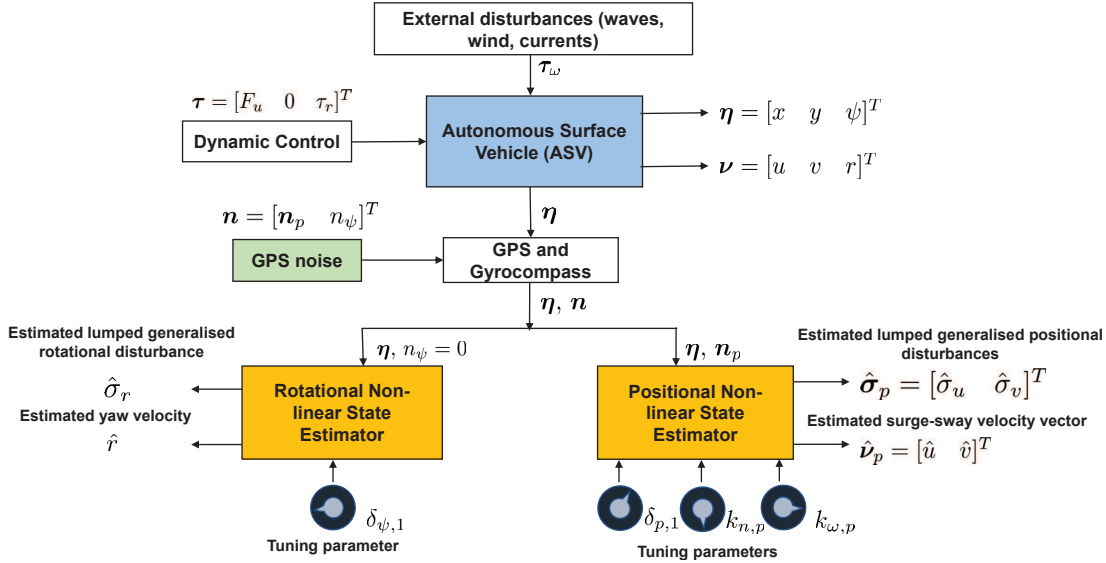}   
	\caption{Block diagram of the complete system}
	\label{figGA}
\end{figure*}

Hence, solving the optimization problem \eqref{eqPositionalOptimizationProblem} with objective function \eqref{eqPositionalOptimizationProblemObjFunction} and constraints \eqref{eqPositionalOptimizationProblemPositiveDefiniteLMI}-\eqref{eqPositionalOptimizationProblemTradeoffLMI} and \eqref{eqPositionalOptimizationProblemPolePlacementLMI} ensures an ISS positional observer with respect to the disturbances $\bm{\omega}_p$ and noises $\bm{n}_p$, with the minimization of the final estimation error bound. The tuning parameters of the positional observer are the scalars $\delta_{p,1}$, $k_{\omega,p}$, and $k_{n,p}$, which module the velocity of convergence, robustness against disturbances, and noise rejection, respectively. For the sake of a better understanding of the available degrees of freedom, Table \ref{tabTuningParam} summarizes the influence of each tuning parameter, which may serve as a good help for the user, while \figurename \ref{figGA} represents a block diagram of the complete proposed observer.

\begin{table*}[!t]
	\caption{INFLUENCE OF THE TUNING PARAMETERS OF THE PROPOSED ESTIMATOR} 
	\label{tabTuningParam}
	\centering
	\begin{tabular}{cccc}
		\bfseries{Parameter} & \bfseries{Low values} & \bfseries{High values} & \bfseries{Notes} \\
		\hline
		$\delta_{\psi,1}$ & Higher steady-state error \&  & Lower steady-state error \& & No noise assumed\\
		& lower velocity of convergence & higher velocity of convergence  \\
		\hline
		$\delta_{p,1}$ & Lower steady-state error & Higher velocity of convergence & With noise \\
		\hline
		$k_{\omega,p}$ & Lower disturbance rejection & Higher disturbance rejection & For fixed $k_{n,p}$ \\
		\hline
		$k_{n,p}$ & Higher noise sensitivity & Lower noise sensitivity & For fixed $k_{\omega,p}$ \\
		\hline
	\end{tabular}
\end{table*}

\begin{remark} \label{remarkPositionalFeasibility}
	The optimization problem \eqref{eqPositionalOptimizationProblem}, including the modification in Remark \ref{remarkPositionalPolePlacement}, is feasible when $r_{\min} = r_{\max} = 0$, since the pair ($\bm{A}_{0,p}$, $\bm{C}_{p}$) is observable. The increase of those values, one positive ($r_{\max}$) and the other one negative ($r_{\min}$), will shrink the space of solutions in $\bm{P}_{p}$, hindering the feasibility of the problem. A formal feasibility analysis is left for future work. 
\end{remark}

\section{Simulation results} \label{secSimulation}

This section shows several simulation results that verify the effectiveness and robustness of the proposed estimation strategy and compare it to similar approaches. The CyberShip II (a 1:70 scale replica of a supply ship for the North Sea) has been used as the test-bed; its geometric and hydrodynamic identified parameters have been obtained from \cite{skjetne2004nonlinear} and are summarised in Table \ref{tabVesselParameters}. Notice that not all these parameters are required to be accurately known by the observer, since only estimates of the parameters included in the inertial matrix $\bm{M}$ described in \eqref{eqMatrixM} are required to implement the observer, as shown in Section \ref{secObserverDesign}, but they are necessary to simulate the CyberShip II performance.

\begin{table}[!ht]
	\renewcommand{\arraystretch}{1.3}
	\caption{PARAMETERS OF THE CYBERSHIP II \cite{skjetne2004nonlinear}} 
	\label{tabVesselParameters}
	\centering
	\begin{tabular}{>{\centering}m{0.05\textwidth} >{\centering}m{0.14\textwidth} |>{\centering}m{0.05\textwidth} >{\centering\arraybackslash}m{0.14\textwidth}}
		\bfseries{Parameter} & \bfseries{Value} & \bfseries{Parameter} & \bfseries{Value} \\
		\hline
		$m$ & 23.8 kg & $Y_{v|v|}$ & -36.47287 kg m\textsuperscript{-1} \\
		$I_z$ & 1.76 kg m\textsuperscript{2} & $Y_{v|r|}$ & -0.805 kg \\ 
		$x_g$ & 0.046 m & $Y_r$ & -7.25 kg m s \textsuperscript{-1} \\
		$X_{\dot{u}}$ & -2 kg & $Y_{r|v|}$ & -0.845 kg \\
		$Y_{\dot{v}}$ & -10 kg & $Y_{r|r|}$ & -3.45 kg m \\
		$Y_{\dot{r}}$ & 0 kg m & $N_v$ & 0.0313 kg m s\textsuperscript{-1} \\
		$N_{\dot{v}}$ & 0 kg m & $N_{v|v|}$ & 3.95645 kg \\
		$N_{\dot{r}}$ & -1 kg m\textsuperscript{2} & $N_{v|r|}$ & 0.13 kg m \\
		$X_u$ & -0.72253 kg s\textsuperscript{-1} & $N_r$ & -1.9 kg m\textsuperscript{2} s\textsuperscript{-1} \\
		$X_{u|u|}$ & -1.32742 kg m\textsuperscript{-1} & $N_{r|v|}$ & 0.08 kg m \\
		$Y_v$ & -0.88965 kg s\textsuperscript{-1} & $N_{r|r|}$ & -0.75 kg m\textsuperscript{2} \\
	\end{tabular}
\end{table}


Since the performance of the estimator does not depend on the control law, as stated in Section \ref{sec:introduction}, open-loop simulations are presented below. In particular, sinusoidal signals with random amplitude, frequency and bias are applied to the control actions (force $F_u$ and torque $\tau_r$), as well as to the external disturbances $F_{w,u}$, $F_{w,v}$, and $\tau_{w,r}$, to generate realistic vessel motion (positive surge, sway lower or equal than surge, yaw achievable for standard propellers). In the next subsections, the initial kinematics and dynamics of the ASV are $\bm{\eta}(t=0) = \left[0 \; 10 \; 0\right]^{T}$ and $\bm{\nu}(t=0) = \left[0 \; 0 \; 0\right]^{T}$, whereas the initial condition for the estimator is $\hat{\bm{\eta}}(t=0) = \hat{\bm{\nu}}(t=0) = \hat{\bm{\sigma}}(t=0) = \left[0 \; 0 \; 0\right]^{T}$.

\subsection{Comparison with Similar Approaches}

The proposed estimator will be compared with the one described in \cite{liu2019state}, which was proposed for noiseless measurements. For the noiseless scenario, the tuning parameters are chosen as $k_{n,p}$ = 0, $k_{\omega,p}$ = 1, thus all the effort is put into disturbance rejection. The considered maximum and minimum values of the yaw velocity are $r_{\max} = -r_{\min}$ = 0.8727~rad~s\textsuperscript{-1}.

Table \ref{tabNoiselessSimulation} includes the standard deviation of the estimation errors for the velocities and disturbances for a complete simulation. The results provided by the observer in \cite{liu2019state} and the one proposed here are compared. The tuning parameters, $\epsilon$ for the observer in \cite{liu2019state} and $\delta_{i,1} \; \; \text{for}\; i=p,\psi$, control the velocity of convergence. 

\begin{table*}[ht]
    \caption{STANDARD DEVIATION OF ESTIMATION ERRORS FOR SEVERAL ESTIMATORS: NOISELESS CASE.} \label{tabNoiselessSimulation}
	\centering
		\begin{tabular}{cccccccc}
			Estimator & Parameter & $e_u$ (m s\textsuperscript{-1}) & $e_v$ (m s\textsuperscript{-1}) & $e_r$ (rad s \textsuperscript{-1})& $e_{\sigma_u}$ (m s\textsuperscript{-2}) & $e_{\sigma_{v}}$ (m s\textsuperscript{-2})& $e_{\sigma_{r}}$ (rad s\textsuperscript{-2}) \\
			\hline
			 \cite{liu2019state} & $\epsilon = 0.1$ & $1.9\cdot10^{-3}$ & $1.8\cdot10^{-3}$ & $2.8\cdot10^{-3}$ & $1.9\cdot10^{-2}$ & $1.8\cdot10^{-2}$ & $2.8\cdot10^{-2}$  \\
			& $\epsilon=0.006$ & $7.2\cdot10^{-6}$ & $6.9\cdot10^{-6}$ & $1.0\cdot10^{-5}$ & $1.6\cdot10^{-3}$ & $1.1\cdot10^{-3}$ & $1.7\cdot10^{-3}$ \\ 
			\hline
			Proposed & $\delta_{\psi,1}=0.05$ & \multirow{2}{*}{$3.9\cdot10^{-6}$}  & \multirow{2}{*}{$3.7\cdot10^{-6}$} & \multirow{2}{*}{$8.9\cdot10^{-6}$} & \multirow{2}{*}{$7.3\cdot10^{-3}$} & \multirow{2}{*}{$6.9\cdot10^{-3}$} & \multirow{2}{*}{$8.8\cdot10^{-3}$} \\
			& $\delta_{p,1}=0.05$ & & & & & \\
			& $\delta_{\psi,1}=300$ & 
			\multirow{2}{*}{$3.9\cdot10^{-6}$} & \multirow{2}{*}{$3.7\cdot10^{-6}$} & \multirow{2}{*}{$\bm{3.1\cdot10^{-7}}$} & 
			\multirow{2}{*}{$7.2\cdot10^{-3}$}  & \multirow{2}{*}{$6.8\cdot10^{-3}$}  & \multirow{2}{*}{\bm{$3.0\cdot10^{-4}$}} \\
			& $\delta_{p,1}=0.05$ & & & & & \\
			& $\delta_{\psi,1}=0.05$ & 
			\multirow{2}{*}{$\bm{1.9\cdot10^{-7}}$}
			& \multirow{2}{*}{$\bm{1.8\cdot10^{-7}}$} & \multirow{2}{*}{$8.9\cdot10^{-6}$} &
			\multirow{2}{*}{$\bm{1.1\cdot10^{-3}}$} & \multirow{2}{*}{$\bm{1.9\cdot10^{-4}}$} & \multirow{2}{*}{$8.8\cdot10^{-3}$}\\ 
			& $\delta_{p,1}=400$ & & & & & \\
			\hline
		\end{tabular}
\end{table*}

Several conclusions may be drawn from the simulations included in Table \ref{tabNoiselessSimulation}. First of all, for the noiseless case, both $\epsilon$ and the pair $\delta_{i,1}$ can be tuned to reduce the error (approaching zero for the ideal noiseless case). However, the proposed algorithm allows to weight the rotational and positional components differently and thus provides the user with more freedom. This fact is illustrated by the values in bold in Table \ref{tabNoiselessSimulation}. 

When the measurements are affected by noises, the results are more interesting. The tuning parameters have been chosen as $k_{n,p} = k_{\omega,p} = 1$. As before, $r_{\max} = -r_{\min}$ = 0.8727~rad~s\textsuperscript{-1}. Uniformly distributed random noises are considered in the position measurement, within $\pm 0.2$ m and frequency range between 0.1 and 0.5 Hz\footnote{State-of-the-art smoother algorithms can be employed to filter the GPS signal, such as \cite{liu2010two}.}. The numerical results are included in Table \ref{tabNoiseSimulation}. Only the positional variables are included, since no noise is assumed to affect the rotational dynamics.

Notice that, in the presence of noises, increasing the velocity of convergence (by means of a reduction of $\epsilon$ or an increase of $\delta_{p,1}$) does not imply better performance as before. While the estimation errors on the surge and sway may be reduced, the errors on the lumped disturbances are increased and the whole state vector estimation turns out to be worse. This is the expected behaviour from \eqref{eqPositionalErrorBound}, as faster dynamics will imply larger bounds for the final error and thus worse steady-state performance. Again, by suitable tuning of the observer parameters, the proposed estimator is able to reduce to a certain extent the estimation errors compared to the observer presented in \cite{liu2019state}. 

\begin{table*}[!ht]
    \renewcommand{\arraystretch}{1}
    \caption{STANDARD DEVIATION OF ESTIMATION ERRORS FOR SEVERAL ESTIMATORS: NOISY CASE} \label{tabNoiseSimulation}
	\centering
		\begin{tabular}{cccccc}
			Estimator & Parameter & $e_u$ (m s\textsuperscript{-1}) & $e_v$ (m s\textsuperscript{-1}) & $e_{\sigma_u}$ (m s\textsuperscript{-2}) & $e_{\sigma_{v}}$ (m s\textsuperscript{-2}) \\
			\hline
			\cite{liu2019state} & $\epsilon = 0.2$ & 
			$0.0898$ & $0.0963$ & $0.1028$ & $0.1130$  \\
			& $\epsilon=0.1$ & $0.0730$ & $0.0788$ & $0.1369$ & $0.1498$ \\ 
			\hline
			Proposed & $\delta_{\psi,1}=\delta_{p,1}=1.2$ & $0.0822$ & $0.0890$ & $0.0845$ & $0.0944$ \\
			& $\delta_{\psi,1}=\delta_{p,1}=2.2$ & 
			$0.0713$ & $0.0765$ & 
			$0.0899$ & $0.1002$ \\
			\hline
		\end{tabular}
\end{table*}

\begin{figure}[!ht]
	\includegraphics[width=9.6cm]{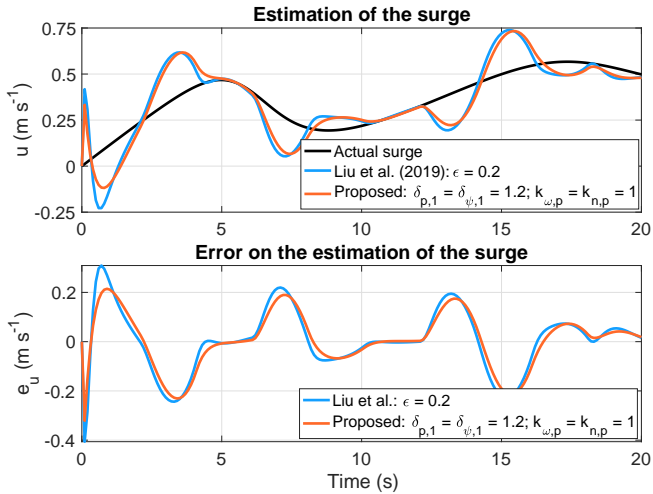}	
	\caption{Estimation performance of the surge $u$ for two different estimators} 
	\label{figEstSurge_Noise}
\end{figure}

\begin{figure}[!ht]
	\includegraphics[width=9.6cm]{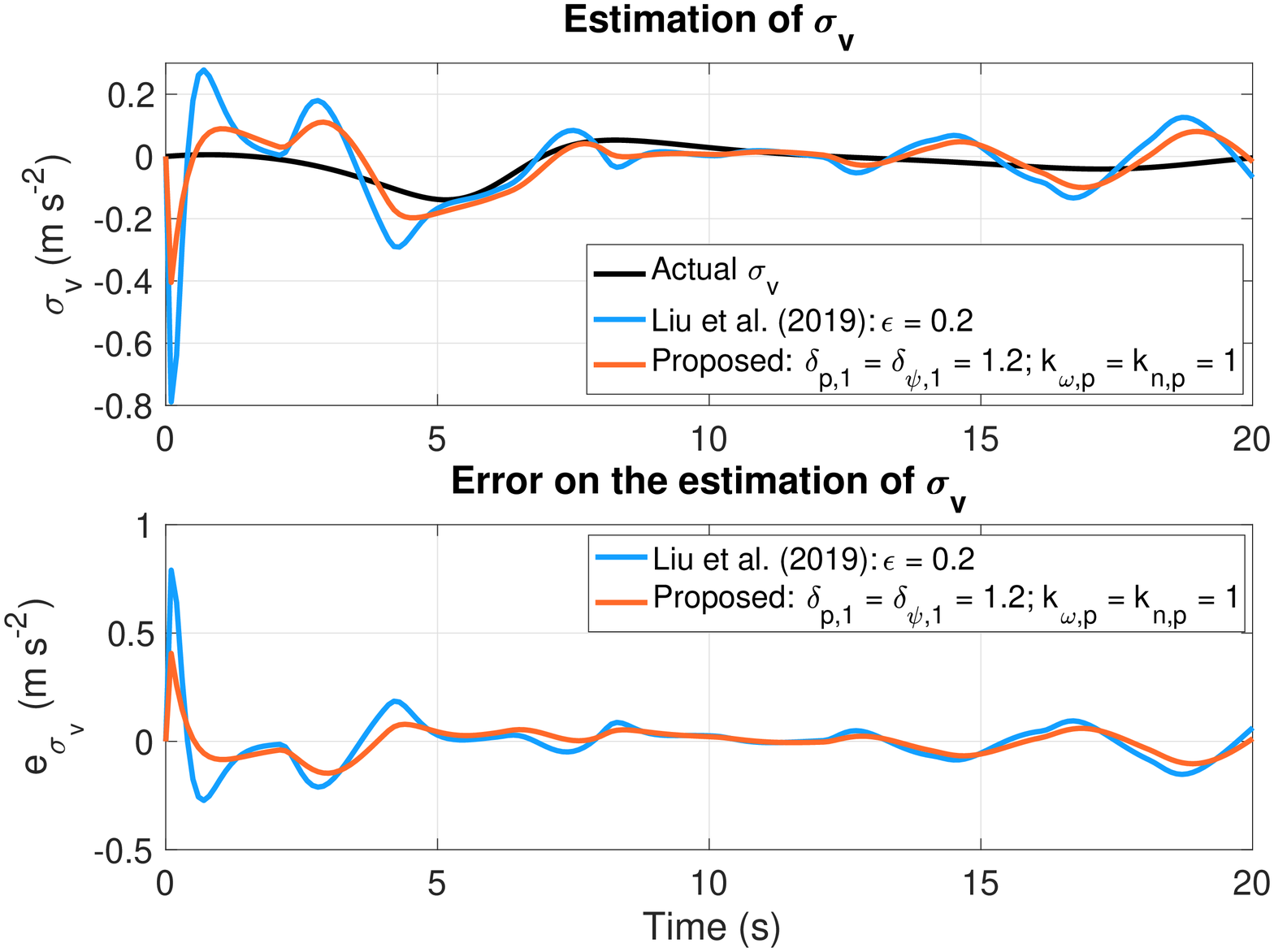}
	\caption{Estimation performance of the sway disturbance $\sigma_v$ for two different estimators}  
	\label{figEstSwayDisturb_Noise}
\end{figure}

Finally, some graphical results are included. \figurename \ref{figEstSurge_Noise}--\ref{figEstSwayDisturb_Noise} depict the estimation of the surge velocity $u$ and the sway-related disturbance $\sigma_v$, respectively, for the best two cases analysed before. These plots illustrate that the proposed observer is able to provide more accurate estimates of both velocities and lumped disturbances.  

\subsection{Tuning Capabilities of the Proposed Observer}

This section intends to show the tuning capabilities that the proposed observer provides to the user. Parameter $\delta_{\psi,1}$ controls the velocity of convergence of the rotational estimator. Regarding the positional observer, parameter $\delta_{p,1}$ modules the velocity of convergence, while parameters $k_{n,p}$ and $k_{\omega,p}$ allow to achieve a user-defined trade-off between noise rejection and robustness against disturbances. 

\begin{figure}[!ht]
	\includegraphics[width=8.8cm]{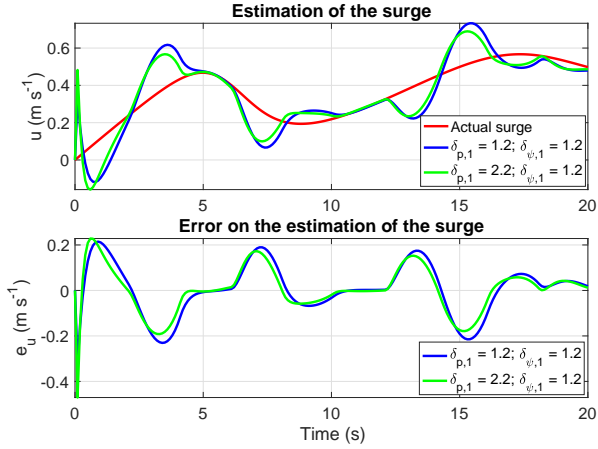}	
	\caption{Estimation performance of the surge $u$ for two different values of $\delta_{p,1}$} 
	\label{figEstSurge_delta1}
\end{figure}

\begin{figure}[!ht]
	\includegraphics[width=8.8cm]{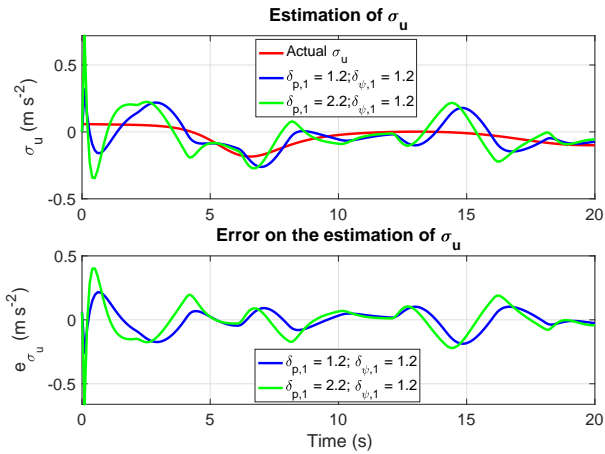}	
	\caption{Estimation performance of the surge disturbance $\sigma_u$ for two different values of $\delta_{p,1}$ } 
	\label{figEstSigmaU_delta1}
\end{figure}

Firstly, an analysis of the influence of $\delta_{p,1}$ for fixed $k_{n,p} = k_{\omega,p} = 1$ is provided. 
\figurename \ref{figEstSurge_delta1}--\ref{figEstSigmaU_delta1} depict the estimation of the surge velocity $u$ and the related disturbance $\sigma_u$, respectively. In the presence of noises, higher values of $\delta_{p,1}$ produce more accurate estimates of the velocities, but the estimates of the lumped disturbances are degraded. 
As theoretically expected, a higher value of $\delta_{p,1}$ leads to a more aggressive estimator, since the observer poles are forced to be on the left of $-\delta_{p,1}$ and the required velocity of convergence is greater. This means that the first derivative of the measured variables (velocities) may benefit from this aggressiveness, but the lumped disturbances, which are actually accelerations and thus the second derivative of the measured variables, turn out to be degraded. Therefore, this tuning parameter $\delta_{p,1}$, in addition to controlling the velocity of convergence of the whole positional state vector (including velocities and lumped disturbances), might act as a way of promoting a more accurate estimation of some variables of $\bm{\chi}_p$, according to the user's aim.  As expressed in Table \ref{tabTuningParam}, high values of $\delta_{p,1}$ lead to higher velocity of convergence, but also to greater steady-state errors of the whole positional state vector $\bm{\chi}_p$, which will be focused on the lumped disturbances rather than the velocities, according to the simulation results shown in \figurename \ref{figEstSurge_delta1}--\ref{figEstSigmaU_delta1}.  

\begin{figure}[!ht]
	\includegraphics[width=8.8cm]{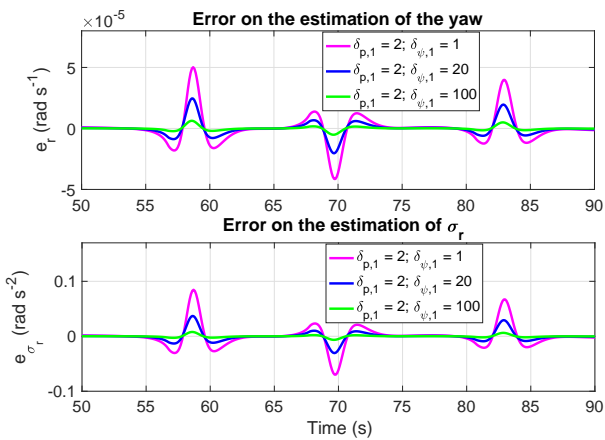}	
	\caption{Estimation performance of the yaw $r$ and the related lumped disturbance $\sigma_r$ for three different values of $\delta_{\psi,1}$}  
	\label{figEstYaw_delta1}
\end{figure}

In the case of the rotational variables $r$ and $\sigma_r$, which are assumed not to be influenced by noise, increasing $\delta_{\psi,1}$ only produces positive effects, as illustrated in Table \ref{tabNoiselessSimulation} of the previous section. An illustrative simulation is shown in \figurename \ref{figEstYaw_delta1}.

\begin{figure}[!ht]
	\includegraphics[width=8.8cm]{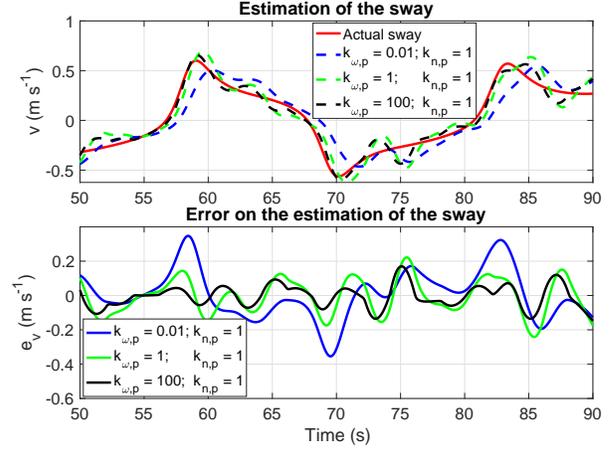}	
	\caption{Estimation performance of the sway $v$ for different values of $k_{\omega,p}$, considering $k_{n,p}$ = 1} 
	\label{figEstSway_kw}
\end{figure}

Finally, this section analyses the influence of the weighting parameters $k_{\omega,p}$ and $k_{n,p}$ on the performance of the estimator. To do this, one of them will be set and the other parameter will be modified. The values of $\delta_{p,1}$ and $\delta_{\psi,1}$ are set to zero, thus the poles of the closed loop dynamics can be located anywhere in the left-half plane. For the analysis, the estimation of the sway velocity $v$ is considered.
\figurename \ref{figEstSway_kw} presents several simulations for different values of those parameters. For the sake of better understanding of the figures, it is worth mentioning that the internal disturbances due to the non-linear behaviour of the ASV are low-frequency signals, while the considered GPS noises are medium-frequency signals.

On the one hand, one can easily see that the higher the ratio ${k_{\omega,p}}/{k_{n,p}}$ is, the more robust is the estimator to the disturbances in the sway. The best results are obtained with $k_{\omega,p} = 100$ and $k_{n,p} = 1$, as appreciated in \figurename \ref{figEstSway_kw}. On the other hand, if the ratio ${k_{\omega,p}}/{k_{n,p}}$ is small, the noise rejection capabilities of the observer are remarked and a more filtered estimation of the sway is produced, thus highlighting the trade-off between disturbance rejection and noise sensitivity provided by the $k_{\omega,p}$ and $k_{n,p}$ tuning parameters.

\section{Conclusions and future work} \label{secConclusions}


In this work, a robust non-linear state estimator for ASVs has been presented, which allows recovering the unmeasured velocities and estimating the lumped generalised disturbances, considering that only the vessel position and orientation are measured, while the position measurement may be affected by bounded noise. The estimation of the lumped disturbances is a key issue to address motion control, since they lump together external non-measurable disturbances, non-linearities, and unmodelled dynamics. The vessel dynamics are decoupled between positional and rotational dynamics, which reveals a cascade structure where the rotational dynamics are independent and the positional ones are coupled with the former ones through the orientation. The observer design benefits from this issue by providing a set of meaningful parameters which allow the user to set the velocity of convergence of the positional and rotational estimates independently, as well as achieve a user-defined trade-off between noise rejection and robustness against disturbances in the case of the positional estimator. The simulation results show that the proposed observer outperforms previous similar approaches in the literature when both noiseless and noisy position measurements are considered. 

Future work will consider the influence of bounded noise also affecting the orientation measurement and an analysis of the impact of the frequency of noises and disturbances on the performance of the estimation. Moreover, the experimental application of the proposed observer to a low-cost in-house-designed ASV is intended in the near future.

\bibliographystyle{IEEEtran}
\bibliography{IEEEabrv,bibliography} 

\begin{thebibliography}{10}
\providecommand{\url}[1]{#1}
\csname url@samestyle\endcsname
\providecommand{\newblock}{\relax}
\providecommand{\bibinfo}[2]{#2}
\providecommand{\BIBentrySTDinterwordspacing}{\spaceskip=0pt\relax}
\providecommand{\BIBentryALTinterwordstretchfactor}{4}
\providecommand{\BIBentryALTinterwordspacing}{\spaceskip=\fontdimen2\font plus
\BIBentryALTinterwordstretchfactor\fontdimen3\font minus
  \fontdimen4\font\relax}
\providecommand{\BIBforeignlanguage}[2]{{%
\expandafter\ifx\csname l@#1\endcsname\relax
\typeout{** WARNING: IEEEtran.bst: No hyphenation pattern has been}%
\typeout{** loaded for the language `#1'. Using the pattern for}%
\typeout{** the default language instead.}%
\else
\language=\csname l@#1\endcsname
\fi
#2}}
\providecommand{\BIBdecl}{\relax}
\BIBdecl

\bibitem{liu2016unmanned}
Z.~Liu, Y.~Zhang, X.~Yu, and C.~Yuan, ``Unmanned surface vehicles: An overview
  of developments and challenges,'' \emph{Annu. Rev. Control}, vol.~41, pp. 71
  -- 93, 2016.

\bibitem{wynn2014autonomous}
R.~B. Wynn, V.~A. Huvenne, T.~P.~L. Bas, B.~J. Murton, D.~P. Connelly, B.~J.
  Bett, H.~A. Ruhl, K.~J. Morris, J.~Peakall, D.~R. Parsons, E.~J. Sumner,
  S.~E. Darby, R.~M. Dorrell, and J.~E. Hunt, ``{Autonomous Underwater Vehicles
  (AUVs): Their past, present and future contributions to the advancement of
  marine geoscience},'' \emph{Mar. Geol.}, vol. 352, pp. 451 -- 468, 2014.

\bibitem{zhang2015future}
F.~{Zhang}, G.~{Marani}, R.~N. {Smith}, and H.~T. {Choi}, ``Future trends in
  marine robotics,'' \emph{IEEE Robot. Autom. Mag.}, vol.~22, no.~1, pp.
  14--122, 2015.

\bibitem{miao2017compound}
J.~Miao, S.~Wang, M.~M. Tomovic, and Z.~Zhao, ``Compound line-of-sight
  nonlinear path following control of underactuated marine vehicles exposed to
  wind, waves, and ocean currents,'' \emph{Nonlinear Dyn.}, vol.~89, no.~4, pp.
  2441--2459, 2017.

\bibitem{wang2019fuzzy}
N.~Wang, Z.~Sun, J.~Yin, Z.~Zou, and S.-F. Su, ``Fuzzy unknown observer-based
  robust adaptive path following control of underactuated surface vehicles
  subject to multiple unknowns,'' \emph{Ocean Eng.}, vol. 176, pp. 57 -- 64,
  2019.

\bibitem{xia2019improved}
Y.~Xia, K.~Xu, Y.~Li, G.~Xu, and X.~Xiang, ``{Improved line-of-sight trajectory
  tracking control of under-actuated AUV subjects to ocean currents and input
  saturation},'' \emph{Ocean Eng.}, vol. 174, pp. 14 -- 30, 2019.

\bibitem{breivik2010topics}
M.~Breivik, ``Topics in guided motion control of marine vehicles,'' Ph.D.
  dissertation, Norwegian University of Science and Technology, 2010.

\bibitem{chen2017sliding}
M.~Chen, S.-D. Chen, and Q.-X. Wu, ``{Sliding mode disturbance observer-based
  adaptive control for uncertain MIMO nonlinear systems with dead-zone},''
  \emph{Int. J. Adapt. Control Signal Process.}, vol.~31, no.~7, pp.
  1003--1018, 2017.

\bibitem{CUI201645}
R.~Cui, X.~Zhang, and D.~Cui, ``Adaptive sliding-mode attitude control for
  autonomous underwater vehicles with input nonlinearities,'' \emph{Ocean
  Eng.}, vol. 123, pp. 45 -- 54, 2016.

\bibitem{lekkas2014integral}
A.~M. {Lekkas} and T.~I. {Fossen}, ``{Integral LOS path following for curved
  paths based on a monotone cubic hermite spline parametrization},'' \emph{IEEE
  Trans. Control Syst. Technol.}, vol.~22, no.~6, pp. 2287--2301, 2014.

\bibitem{liu2019state}
L.~Liu, D.~Wang, and Z.~Peng, ``State recovery and disturbance estimation of
  unmanned surface vehicles based on nonlinear extended state observers,''
  \emph{Ocean Eng.}, vol. 171, pp. 625--632, 2019.

\bibitem{skjetne2005adaptive}
R.~Skjetne, T.~I. Fossen, and P.~V. Kokotovi{\'c}, ``Adaptive maneuvering, with
  experiments, for a model ship in a marine control laboratory,''
  \emph{Autom.}, vol.~41, no.~2, pp. 289 -- 298, 2005.

\bibitem{yin2017tracking}
S.~{Yin} and B.~{Xiao}, ``{Tracking Control of Surface Ships With Disturbance
  and Uncertainties Rejection Capability},'' \emph{IEEE/ASME Trans.
  Mechatron.}, vol.~22, no.~3, pp. 1154--1162, 2017.

\bibitem{peng2020output}
Z.~{Peng}, L.~{Liu}, and J.~{Wang}, ``{Output-Feedback Flocking Control of
  Multiple Autonomous Surface Vehicles Based on Data-Driven Adaptive Extended
  State Observers},'' \emph{IEEE Trans. Cybern.}, pp. 1--12, 2020.

\bibitem{manley2008unmanned}
J.~E. Manley, ``Unmanned surface vehicles, 15 years of development,'' in
  \emph{OCEANS 2008}.\hskip 1em plus 0.5em minus 0.4em\relax IEEE, 2008, pp.
  1--4.

\bibitem{dai2018adaptive}
S.-L. Dai, S.~He, M.~Wang, and C.~Yuan, ``Adaptive neural control of
  underactuated surface vessels with prescribed performance guarantees,''
  \emph{IEEE Trans. Neural Netw. Learn. Syst.}, vol.~30, no.~12, pp.
  3686--3698, 2018.

\bibitem{peng2019path}
Z.~{Peng}, J.~{Wang}, and Q.~{Han}, ``{Path-Following Control of Autonomous
  Underwater Vehicles Subject to Velocity and Input Constraints via
  Neurodynamic Optimization},'' \emph{IEEE Trans. Ind. Electron.}, vol.~66,
  no.~11, pp. 8724--8732, 2019.

\bibitem{do2006underactuated}
K.~D. Do and J.~Pan, ``Underactuated ships follow smooth paths with integral
  actions and without velocity measurements for feedback: theory and
  experiments,'' \emph{IEEE Trans. Control Syst. Tech.}, vol.~14, no.~2, pp.
  308--322, 2006.

\bibitem{feemster2011comprehensive}
M.~G. Feemster and J.~M. Esposito, ``Comprehensive framework for tracking
  control and thrust allocation for a highly overactuated autonomous surface
  vessel,'' \emph{J. Field Robot.}, vol.~28, no.~1, pp. 80--100, 2011.

\bibitem{pan2013efficient}
C.-Z. Pan, X.-Z. Lai, S.~X. Yang, and M.~Wu, ``An efficient neural network
  approach to tracking control of an autonomous surface vehicle with unknown
  dynamics,'' \emph{Expert Syst. Appl.}, vol.~40, no.~5, pp. 1629--1635, 2013.

\bibitem{chen2012robust}
M.~Chen, S.~S. Ge, B.~V.~E. How, and Y.~S. Choo, ``Robust adaptive position
  mooring control for marine vessels,'' \emph{IEEE Trans. Control Syst.
  Technol.}, vol.~21, no.~2, pp. 395--409, 2012.

\bibitem{tee2006control}
K.~P. Tee and S.~S. Ge, ``Control of fully actuated ocean surface vessels using
  a class of feedforward approximators,'' \emph{IEEE Trans. Control Syst.
  Technol.}, vol.~14, no.~4, pp. 750--756, 2006.

\bibitem{ZHANG20111430}
L.-J. Zhang, H.-M. Jia, and X.~Qi, ``{NNFFC-adaptive output feedback trajectory
  tracking control for a surface ship at high speed},'' \emph{Ocean Eng.},
  vol.~38, no.~13, pp. 1430 -- 1438, 2011.

\bibitem{motwani2013interval}
A.~Motwani, S.~Sharma, R.~Sutton, and P.~Culverhouse, ``{Interval Kalman
  filtering in navigation system design for an uninhabited surface vehicle},''
  \emph{J. Navig.}, vol.~66, no.~5, pp. 639--652, 2013.

\bibitem{peng2009adaptive}
Y.~Peng, J.~Han, and Q.~Huang, ``{Adaptive UKF based tracking control for
  unmanned trimaran vehicles},'' \emph{Int. J. Innov. Comput., Inf. Control},
  vol.~5, no.~10, pp. 3505--3516, 2009.

\bibitem{tran2014tracking}
N.-H. Tran, H.-S. Choi, S.-H. Baek, and H.-Y. Shin, ``Tracking control of an
  unmanned surface vehicle,'' in \emph{AETA 2013: Recent Advances in Electrical
  Engineering and Related Sciences}.\hskip 1em plus 0.5em minus 0.4em\relax
  Springer, 2014, pp. 575--584.

\bibitem{peng2016cooperative}
Z.~Peng, D.~Wang, and J.~Wang, ``{Cooperative Dynamic Positioning of Multiple
  Marine Offshore Vessels: A Modular Design},'' \emph{IEEE/ASME Trans.
  Mechatron.}, vol.~21, no.~3, pp. 1210--1221, 2016.

\bibitem{peng2017distributed}
Z.~{Peng}, J.~{Wang}, and D.~{Wang}, ``Distributed containment maneuvering of
  multiple marine vessels via neurodynamics-based output feedback,'' \emph{IEEE
  Trans. Ind. Electron.}, vol.~64, no.~5, pp. 3831--3839, 2017.

\bibitem{peng2018output}
Z.~{Peng} and J.~{Wang}, ``{Output-Feedback Path-Following Control of
  Autonomous Underwater Vehicles Based on an Extended State Observer and
  Projection Neural Networks},'' \emph{IEEE Trans. Syst, Man, Cybern.: Syst.},
  vol.~48, no.~4, pp. 535--544, 2018.

\bibitem{peng2019output}
Z.~Peng, D.~Wang, T.~Li, and M.~Han, ``Output-feedback cooperative formation
  maneuvering of autonomous surface vehicles with connectivity preservation and
  collision avoidance,'' \emph{IEEE Trans. Cybern.}, vol.~50, no.~6, pp.
  2527--2535, 2019.

\bibitem{gu2019observer}
N.~Gu, D.~Wang, Z.~Peng, and L.~Liu, ``{Observer-Based Finite-Time Control for
  Distributed Path Maneuvering of Underactuated Unmanned Surface Vehicles With
  Collision Avoidance and Connectivity Preservation},'' \emph{IEEE Trans. Syst,
  Man, Cybern.: Syst.}, 2019.

\bibitem{zhang2019fixed}
J.~Zhang, S.~Yu, and Y.~Yan, ``Fixed-time extended state observer-based
  trajectory tracking and point stabilization control for marine surface
  vessels with uncertainties and disturbances,'' \emph{Ocean Eng.}, vol. 186,
  p. 106109, 2019.

\bibitem{yiang2020line}
Y.~{Jiang}, Z.~{Peng}, D.~{Wang}, and C.~L.~P. {Chen}, ``{Line-of-Sight Target
  Enclosing of an Underactuated Autonomous Surface Vehicle With Experiment
  Results},'' \emph{IEEE Trans. Ind. Inform.}, vol.~16, no.~2, pp. 832--841,
  2020.

\bibitem{li2016adaptive}
Y.~Li and S.~Tong, ``Adaptive fuzzy output-feedback stabilization control for a
  class of switched nonstrict-feedback nonlinear systems,'' \emph{IEEE Trans.
  Cybern.}, vol.~47, no.~4, pp. 1007--1016, 2016.

\bibitem{li2016hybrid}
Y.~{Li}, S.~{Tong}, and T.~{Li}, ``{Hybrid Fuzzy Adaptive Output Feedback
  Control Design for Uncertain MIMO Nonlinear Systems With Time-Varying Delays
  and Input Saturation},'' \emph{IEEE Trans. Fuzzy Syst.}, vol.~24, no.~4, pp.
  841--853, 2016.

\bibitem{sontag2008input}
E.~D. Sontag, ``Input to state stability: Basic concepts and results,'' in
  \emph{Nonlinear and optimal control theory}.\hskip 1em plus 0.5em minus
  0.4em\relax Springer, 2008, pp. 163--220.

\bibitem{khalil2002nonlinear}
H.~K. Khalil and J.~W. Grizzle, \emph{Nonlinear systems}.\hskip 1em plus 0.5em
  minus 0.4em\relax Prentice Hall Upper Saddle River, NJ, 2002, vol.~3.

\bibitem{fossen2011handbook}
T.~I. Fossen, \emph{Handbook of marine craft hydrodynamics and motion
  control}.\hskip 1em plus 0.5em minus 0.4em\relax John Wiley \& Sons, United
  Kingdom, 2011.

\bibitem{skjetne2004nonlinear}
R.~Skjetne, {\O}.~N. Smogeli, and T.~I. Fossen, ``{A nonlinear ship
  manoeuvering model: Identification and adaptive control with experiments for
  a model ship},'' \emph{Model. Identif. Control}, vol.~25, pp. 3--27, 2004.

\bibitem{sname1950nomenclature}
{The Society of Naval Architecture and Marine Engineers}, ``Nomenclature for
  treating the motion of a submerged body through a fluid,'' \emph{Tech. Res.
  Bull. No.}, pp. 1--5, 1950.

\bibitem{zhao2015extended}
Z.-L. Zhao and B.-Z. Guo, ``Extended state observer for uncertain lower
  triangular nonlinear systems,'' \emph{Syst. Control Lett.}, vol.~85, pp. 100
  -- 108, 2015.

\bibitem{lindegaard2003acceleration}
K.-P. Lindegaard, ``Acceleration feedback in dynamic positioning,'' Ph.D.
  dissertation, Norwegian University of Science and Technology, 2003.

\bibitem{vasconcelos2010discrete}
J.~F. Vasconcelos, B.~Cardeira, C.~Silvestre, P.~Oliveira, and P.~Batista,
  ``Discrete-time complementary filters for attitude and position estimation:
  Design, analysis and experimental validation,'' \emph{IEEE Trans. Control
  Syst. Technol.}, vol.~19, no.~1, pp. 181--198, 2010.

\bibitem{vasconcelos2011ins}
J.~Vasconcelos, C.~Silvestre, and P.~Oliveira, ``{INS/GPS aided by frequency
  contents of vector observations with application to autonomous surface
  crafts},'' \emph{IEEE J. Ocean. Eng.}, vol.~36, no.~2, pp. 347--363, 2011.

\bibitem{zhang2006schur}
F.~Zhang, \emph{{The Schur complement and its applications}}.\hskip 1em plus
  0.5em minus 0.4em\relax Springer Science \& Business Media, 2006, vol.~4.

\bibitem{SEDUMI}
J.~F. Sturm, ``{Using SeDuMi 1.02, A Matlab toolbox for optimization over
  symmetric cones},'' \emph{Optim. Methods Softw.}, vol.~11, no. 1-4, pp.
  625--653, 1999.

\bibitem{YALMIP}
J.~L{\"{o}}fberg, ``{YALMIP: A toolbox for modeling and optimization in
  MATLAB},'' in \emph{2004 IEEE Int. Conf. Robot. Autom.}\hskip 1em plus 0.5em
  minus 0.4em\relax IEEE, 2004, pp. 284--289.

\bibitem{chilali1999robust}
M.~Chilali, P.~Gahinet, and P.~Apkarian, ``{Robust pole placement in LMI
  regions},'' \emph{IEEE Trans. Autom. Control}, vol.~44, no.~12, pp.
  2257--2270, 1999.

\bibitem{liu2010two}
H.~Liu, S.~Nassar, and N.~El-Sheimy, ``Two-filter smoothing for accurate
  {INS/GPS} land-vehicle navigation in urban centers,'' \emph{IEEE Trans. Veh.
  Technol.}, vol.~59, no.~9, pp. 4256--4267, 2010.

\end{thebibliography}

\end{document}